\documentclass[usenatbib]{mn2e}
\usepackage{graphicx}
\usepackage{times}
\usepackage{epsfig}

\newcommand{\be}{\begin{equation}}
\newcommand{\ee}{\end{equation}}
\newcommand{\bea}{\begin{eqnarray}}
\newcommand{\eea}{\end{eqnarray}}
\newcommand{\nn}{\nonumber}

\begin{document}

\title[Tidal coupling in white dwarf binaries]{Non-dissipative tidal synchronization in accreting binary white dwarf systems}
\author[\'{E}. Racine, E. S. Phinney and P. Arras]{\'{E}tienne Racine,$^1$ E. Sterl Phinney, $^1$ and Phil Arras$^2$ \\ $^1$ Department of physics, mathematics and astronomy, California Institute of Technology, Pasadena, CA 91125 \\ $^2$ Kavli Institute for Theoretical Physics, University of California at Santa Barbara, Santa Barbara, CA 93106}

\maketitle

\begin{abstract}
We study a non-dissipative hydrodynamical mechanism that can stabilize the spin of the accretor in an ultra-compact double white dwarf binary. This novel synchronization mechanism relies on a nonlinear coupling between tides and the uniform (or rigid) rotation mode, which spins down the background star. The essential physics of the synchronization mechanism is summarized as follows. As the compact binary coalesces due to gravitational wave emission, the largest star eventually fills its Roche lobe and accretion starts. The accretor then spins up due to infalling material and eventually reaches a spin frequency where a normal mode of the star is resonantly driven by the gravitational tidal field of the companion. If the resonating mode satisfies a set of specific criteria, which we elucidate in this paper, it exchanges angular momentum with the background star at a rate such that the spin of the accretor locks at this resonant frequency, even though accretion is ongoing. Some of the accreted angular momentum that would otherwise spin up the accretor is fed back to the orbit through this resonant tidal interaction. In this paper we solve analytically a simple dynamical system that captures the essential features of this mechanism. Our analytical study allows us to identify two candidate Rossby modes that may stabilize the spin of an accreting white dwarf in an ultra-compact binary. These two modes are the $l=4,m=2$ and $l=5,m=3$ CFS unstable hybrid {\it r}-modes, which, for an incompressible equation of state, stabilize the spin of the accretor at frequency $2.6\, \omega_{\rm orb}$ and $1.54\, \omega_{\rm orb}$ respectively, where $\omega_{\rm orb}$ is the binary's orbital frequency. For an $n=3/2$ polytrope, the accretor's spin frequency is stabilized at $2.13\, \omega_{\rm orb}$ and $1.41\, \omega_{\rm orb}$ respectively. Since the stabilization mechanism relies on continuously driving a mode at resonance, its lifetime is limited since eventually the mode amplitude saturates due to non-linear mode-mode coupling. Rough estimates of the lifetime of the effect lie from a few orbits to possibly millions of years. We argue that one must include this hydrodynamical stabilization effect to understand stability and survival rate of ultra-compact binaries, which is relevant in predicting the galactic white dwarf gravitational background that LISA will observe. 
\end{abstract}


\section{Motivation, main results and outline}

The collection of $\sim$ 100 to 200 million double white dwarf binaries (WD-WD) populating the Galaxy generates an important gravitational wave background that the planned LISA mission will be sensitive to. Predicting the properties of this gravitational wave background requires accurate population synthesis models, which in turn necessitate precise understanding of dynamics of WD-WD binaries, which we now briefly summarize. 

Upon formation a WD-WD binary will have its constituents separated widely enough so that no mass transfer is occurring. During this detached phase the orbital motion of the binary is well approximated by that of point-particles in Newtonian gravity, supplemented by leading-order tidal coupling and dissipative contributions due to gravitational wave emission, computed from the Burke-Thorne potential. As the binary coalesces, the larger white dwarf will eventually fill its Roche lobe, signaling the onset of mass transfer. A sizable fraction of galactic WD-WD binaries are compact enough so that this mass transfer phase will be reached within a Hubble time. During mass transfer the binary's dynamics is substantially more complicated than during the detached phase, since the accretion rate and orbital parameters are coupled to one another. In particular the question of whether or not WD-WD binaries survive mass transfer for a long time, as opposed to merging shortly after mass transfer begins, is not fully understood at the present time. If indeed such compact binaries are stable, then they will be among the strongest persistent sources contributing to the galactic gravitational wave background for LISA. There are currently at least two observed ultra-compact WD-WD candidate systems, namely RX J0806.3+1527 (\cite{Strohmayer}) and V407 Vul (\cite{Ramsay,StrohmayerII}).  

The problem of stability of accreting white dwarf binaries has been investigated in detail by \cite{marsh}. They show that for mass ratios lying between guaranteed stability and guaranteed instability, the stability of accreting WD-WD binaries is closely related to the strength of the dissipative synchronization torque $\mathcal{T}_{\rm diss}$ that couples the accretor's spin and the orbit, which is assumed to be of the form 
\be\label{disstorque}
\mathcal{T}_{\rm diss} = \frac{1}{\tau_S} I (\Omega - \omega_{\rm orb}),
\ee
where $\tau_S$ is the synchronization timescale, $I$ the accretor's moment of inertia, $\Omega$ the spin frequency of the accretor and $\omega_{\rm orb}$ the binary's orbital frequency. Their result is that for accreting WD-WD binaries to be stable the synchronization timescale must be $\la 10^3$ years. 

The essential contribution of the present paper is to highlight a non-dissipative hydrodynamical mechanism that may affect significantly the evolution of the spin frequency of the accretor, and therefore impact the stability analysis of \cite{marsh}. Specifically we show that resonant tidal excitation of generalized {\it r}-modes\footnote{Generalized {\it r}-modes are predominantly toroidal perturbations whose restoring force is provided by the Coriolis force, which makes them dynamically unimportant in non-rotating stars.} (or Rossby modes) in the accretor may stabilize its spin at a given resonant frequency of order orbital frequency. During this type of hydrodynamical resonance locking (\cite{reslock}) phase, some of the angular momentum carried by accreted material that would contribute to spin up the white dwarf is instead fed back into the orbit, which contributes to make the binary more stable to mass transfer.

\subsection{The tidal stabilization mechanism}

The physical foundation of the stabilization mechanism we analyze is the coupling of the uniform rotation mode with other stellar modes excited by the companion's tidal field. Formally this coupling comes out naturally from second order stellar perturbation theory (\cite{Schenk}), but may be understood schematically as follows. Consider a perturbation of a uniformly rotating star characterized by a fluid Lagrangian displacement vector field $\bmath{\xi}$. The total angular momentum of the perturbed star may be computed as an expansion in $\bmath{\xi}$, which can be formally written as
\be\label{totalJ}
J_{\rm star} = J_0 + J_1[\bmath{\xi}] + J_2[\bmath{\xi},\bmath{\xi}],
\ee      
where $J_0$ is the total angular momentum of the unperturbed star. Perturbation theory of uniformly rotating stars yields that $J_1$ depends only on the uniform rotation mode (\cite{Schenk}), all other stellar modes contributing only in $J_2$. This implies that the first two terms in the right-hand side of (\ref{totalJ}) can be meaningfully combined into a term of the form $I\Omega$, where $I$ is the moment of inertia of a star rotating uniformly at spin frequency $\Omega$, since the angular momentum $J_0 + J_1$ is carried entirely by the uniform rotation mode\footnote{However in situations where the unperturbed star has significant differential rotation, it is not clear if this result carries over as the formalism of \cite{Schenk} has not been yet extended to unperturbed stars with differential rotation.}. Taking a time derivative of (\ref{totalJ}) then yields
\be\label{totaltorque}
\dot{J}_{\rm star} = \frac{d}{dt}(I\Omega) + \frac{d}{dt}J_2[\bmath{\xi},\bmath{\xi}].
\ee
In binary systems one may compute the contribution to the external torque on the star due to non-dissipative tidal coupling\footnote{Here by "non-dissipative tidal coupling" we simply mean the coupling between the mass multipole moments of the accreting star induced by the fluid perturbation and the Newtonian gravitational tidal field of the companion. In this paper we omit dissipative tidal coupling for the purpose of computing the tidal response of the accretor. We do however mention it when discussing stability issues later in the introduction.} as an expansion in $\bmath{\xi}$ as well, using either the binary's equation of motion as done in Appendix \ref{tidalcouplingtorque}, or equivalently the tidal interaction potential. The total torque on the accretor is then the sum of this tidal torque $-\dot{J}_{\rm tidal}$ and the accretion-induced torque $-\dot{J}_{\rm acc}$\footnote{We introduce minus signs here to follow the convention used in the body of the paper, which defines $\dot{J}_{\rm tidal}$ and $\dot{J}_{\rm acc}$ as rates of change of {\it orbital} angular momentum due to tidal interactions and accretion respectively. This implies in particular that $\dot{J}_{\rm acc}$ must be negative.}, leading to the following angular momentum conservation equation
\be\label{spinevolintro}
\frac{d}{dt}(I\Omega) + \dot{J}_2[\bmath{\xi},\bmath{\xi}] = -(\dot{J}_{\rm tidal} + \dot{J}_{\rm acc}).
\ee
Expanding the lagrangian displacement $\bmath{\xi}(\bmath{x},t)$ in terms of normal modes $\bmath{\xi}_\alpha(\bmath{x})$\footnote{In this work the mode eigenfunctions have dimensions of length, so the mode amplitude coefficients $c_\alpha$ are dimensionless. The normalization condition on the eigenfunctions we use is detailed in Appendix \ref{modeidentities}.} as
\be
\bmath{\xi}(\bmath{x},t) = \sum_{\alpha} c_\alpha(t) \bmath{\xi}_\alpha(\bmath{x}) + \,\,{\rm c.c.},
\ee
one finds that $-\dot{J}_{\rm tidal}$ does not equal $\dot{J}_2[\bmath{\xi},\bmath{\xi}]$, the difference between both terms describing how the uniform rotation mode couples to other stellar modes. This coupling however should not depend on whether the perturbed star is part of a binary system or not. Thus there should exist a derivation of this coupling contained entirely within the framework of stellar perturbation theory. This derivation is indeed given in \cite{Schenk} and yields the same answer as Eq.(\ref{spinevolintro}) when specialized to a binary system. 

Now in studies of accreting binary systems it is generally assumed that the difference between the terms $-\dot{J}_{\rm tidal}$ and $\dot{J}_2$ in (\ref{spinevolintro}) is negligible compared to the accretion-induced torque $\dot{J}_{\rm acc}$. In situations where the tidal response of the accretor's normal modes to the gravitational field of its companion lies away from any resonance, this yields a very good approximation. However as the accretor accumulates material and spins up, it becomes possible to sweep through tidal resonances, in which case the back-reaction of the resonating mode on spin frequency evolution may not be neglected. The mode amplitude grows on a short timescale and the difference between $\dot{J}_2$ and $-\dot{J}_{\rm tidal}$ may become comparable in magnitude to $\dot{J}_{\rm acc}$. As we show in this paper, for a simple model where a single mode dominates the tidal response of the star, Eq.(\ref{spinevolintro}) assumes the form
\be\label{freqevolintro}
\frac{d}{dt}{\Omega} = -\frac{1}{I}\dot{J}_{\rm acc} - \bar{\nu}_\alpha \frac{d}{dt} |c_\alpha|^2,
\ee
where $c_\alpha$ is the amplitude of the resonating mode and $\bar{\nu}_\alpha$ is a parameter describing the coupling between the resonating mode and uniform rotation. This parameter typically scales as $\bar{\nu}_\alpha \sim \omega_\alpha$, $\omega_\alpha$ being the mode eigenfrequency. In the context of a binary system one may also interpret the second term in the right-hand side of (\ref{freqevolintro}) as the difference between the external tidal torque due to the companion and the angular momentum carried by the resonant mode. If $\bar{\nu}_\alpha$ is positive and large enough, it may then be possible that when mode amplitude $c_\alpha$ grows rapidly due to resonant excitation, the accretion-induced torque $\dot{J}_{\rm acc}$ is balanced by the hydrodynamical back-reaction term, resulting in a spin that is nearly constant in time. The value of this quasi-static spin frequency is determined by the eigenfrequency of the resonant mode. 

Let us again emphasize that this spin stabilization mechanism is entirely hydrodynamical since it relies solely on tidal excitation of normal modes of a perfect fluid. No internal damping is needed to stabilize the spin in this scenario. However, as opposed to dissipative synchronization, the lifetime of this mechanism is limited since it requires continuously driving a mode near resonance. Eventually the mode amplitude will grow large enough so that non-linear mode-mode coupling cannot be neglected. Generically mode-mode coupling sets a saturation amplitude beyond which it is not possible to drive the mode effectively. The lifetime of our stabilization mechanism therefore depends crucially on this saturation amplitude which, for resonant tidal excitation of modes, is not well known. A computation of this saturation amplitude, as well as understanding the evolution of the binary when the mode saturates, likely requires simulating a large network of coupled modes following the work of \cite{jb1,jb2}, who studied the problem of growth of {\it r}-modes unstable to gravitational radiation reaction. As this is a very complicated dynamical system, we shall remain cautious and not speculate any further on the fate of the binary after saturation. 

\subsection{Overview of the results}

The central result of this paper is the proof that there exists modes in a rotating star whose resonant tidal excitation can stabilize the spin of the accretor in ultra-compact binary white dwarf systems. This result is derived following a number of steps. First we solve analytically the coupled system of equations describing the time evolution of the spin frequency of the accretor $\Omega$  and the mode amplitude $c_\alpha$. This system may be written as
\be
\dot{\Omega} = -\frac{1}{I}\dot{J}_{\rm acc} - \bar{\nu}_\alpha \frac{d}{dt} |c_\alpha|^2, \label{systemintro}
\ee
\be
\dot{c}_\alpha + i\omega_\alpha c_\alpha = \mathcal{F}_\alpha \exp\left[-im_\alpha\left(\phi_{\rm orb} - \int \Omega(t^\prime) dt^\prime \right)\right], \label{modeampevolintro}
\ee
where $\mathcal{F}_\alpha$ is the overlap of the mode eigenfunction with the external tidal field and $\phi_{\rm orb}$ is the binary's orbital phase. To solve system (\ref{systemintro})-(\ref{modeampevolintro}) we make the following three simplifying approximations: (i) we assume the accretion-induced torque is constant, (ii) we neglect the back-reaction of the modes on orbital evolution for the purpose of solving (\ref{modeampevolintro}) and (iii) we assume the accretor is slowly rotating, i.e. $\Omega \ll \sqrt{M/R^3}$. [Throughout we employ geometric units $G=c=1$.] In the case of a mode which stabilizes the spin frequency, the analytic solution for the spin frequency during resonance locking may be written as
\be\label{freqsolintro}
\Omega(t) = \Omega_{\rm res} - \sqrt{\frac{I\bar{\nu}_\alpha|\mathcal{F}_\alpha|^2}{\lambda_\alpha |\dot{J}_{\rm acc}|\, (t - t_0)}},
\ee
where $\Omega_{\rm res}$ is the spin frequency at which mode $\alpha$ resonates, $\lambda_\alpha$ is a dimensionless number of order unity and time $t_0$ is defined as the time when the mode enters resonance regime. 

Once system (\ref{systemintro}) is solved, we investigate whether or not solution (\ref{freqsolintro}) is physically realized by some set of normal modes of the accretor.  We show that solution (\ref{freqsolintro}) is only valid for resonantly excited modes whose back-reaction parameter $\bar{\nu}_\alpha$ and tidal coupling strength $\mathcal{F}_\alpha$ satisfy the following bound
\be\label{boundintro}
\bar{\nu}_{\alpha} |\mathcal{F}_\alpha|^2 \ga 0.4 \left(\frac{|\dot{J}_{\rm acc}|}{I}\right)^{3/2}.
\ee
In order to stabilize the spin of the accretor at frequency of order orbital frequency, one needs to focus attention to modes with low corotating-frame eigenfrequencies, as the stabilization frequency $\Omega_{\rm res}$ is related to orbital frequency $\omega_{\rm orb}$ by
\be\label{omegaresdef}
\Omega_{\rm res} = \omega_{\rm orb} - \frac{\omega_\alpha}{m_\alpha}.
\ee
The ideal candidate modes are generalized (or hybrid) {\it r}-modes (Rossby modes) (\cite{bryan,lindblom}), since their eigenfrequencies scale as $\omega_\alpha \sim \Omega$. Therefore they are very low frequency modes in slowly rotating stars, compared to, say, {\it f} modes. We then show that in ultra-compact binary white dwarf systems the following generalized {\it r}-modes are potential candidates for stabilizing the spin of the accreting star:

\begin{enumerate}

\item The $l=4,m=2$ {\it r}-mode with eigenfrequency $\omega_{42} = -1.232\, \Omega$. This mode stabilizes the spin at frequency $\Omega_{{\rm res}, 42} = 2.60\, \omega_{\rm orb}$. 

\item The $l=5,m=3$ {\it r}-mode with eigenfrequency $\omega_{53} = -1.053 \,\Omega$. This mode stabilizes the spin at frequency $\Omega_{{\rm res}, 53} = 1.54 \,\omega_{\rm orb}$. 

\end{enumerate}
There exist other candidate modes with polar quantum number $l = 6$ but they satisfy bound (\ref{boundintro}) only marginally. 

As mentioned previously Eq.(\ref{freqsolintro}) is valid only up to a certain time $t_{\rm max}$ when the resonant mode amplitude becomes large enough so that non-linear hydrodynamical effects (mode-mode coupling) saturates its growth. In that regime the simple model we analyze here does not describe correctly the tidal response of the star. As mentioned previously the fate of binary in the saturation regime remains an open problem. Denoting the saturation mode amplitude as $c_{\rm sat}$, the lifetime\footnote{By lifetime we here mean the time interval during which the tidal response of star and the distribution of angular momentum is correctly described by the model studied in this paper.}  $T_{\rm lif}$ of the hydrodynamical stabilizing mechanism is then found to be
\be
T_{\rm lif} \sim |c_{\rm sat}|^2 \frac{I \bar{\nu}_\alpha}{|\dot{J}_{\rm acc}|}.
\ee
The current uncertainty in the saturation amplitude and different choices of orbital parameters together give estimates for the lifetime of the mechanism ranging from a few orbits to hundreds, possibly millions of years. We also investigate how our results change when the stellar model is an $n = 3/2$ polytrope. In that case we find that backreaction of the $l=5,m=3$ CFS unstable {\it r}-mode may or may not be strong enough to stabilize the spin, depending on accretion rate and compactness of the binary. If it is not strong enough, then the $l=4,m=2$ CFS unstable mode will be able to stabilize the spin in the case of an $n=3/2$ polytrope.

\subsection{Implications of the stabilization mechanism on binary stability and observations}  

Before going ahead with the detailed derivation of the results presented above, we will discuss briefly some observational consequences of our tidal stabilization mechanism. More specifically we take a look at how the stability analysis of \cite{marsh} is modified when the binary is in the resonance locking regime. During resonance locking the spin frequency of the accretor is essentially constant, so Eq.(\ref{freqevolintro}) yields
\be\label{reslockmodes}
\frac{d}{dt}|c_\alpha|^2 = - \frac{\dot{J}_{\rm acc}}{I\bar{\nu}_\alpha}.
\ee
The evolution equation of the orbital angular momentum is given by
\be\label{Jconsintro}
\dot{J}_{\rm orb} = \dot{J}_{\rm GW} + \dot{J}_{\rm acc} + \dot{J}_{\rm tidal,diss} - m_\alpha b_\alpha \frac{d}{dt}|c_\alpha|^2, 
\ee
where $\dot{J}_{\rm GW}$ is the angular momentum loss due to gravitational waves, $\dot{J}_{\rm tidal, diss}$ is the dissipative tidal torque usually modeled as in Eq.(\ref{disstorque}) and the last term of the right-hand side is due to modal tidal coupling [see Appendix \ref{tidalcouplingtorque}]. The number $b_\alpha$, sometimes called wave action, is mode energy at unit amplitude divided by corotating-frame mode frequency. Substituting (\ref{reslockmodes}) into (\ref{Jconsintro}) yields
\be\label{JconsintroII}
\dot{J}_{\rm orb} = \dot{J}_{\rm GW} + (1-x_\alpha)\dot{J}_{\rm acc} + \dot{J}_{\rm tidal,diss}, 
\ee
where 
\be
x_\alpha = - \frac{m_\alpha b_\alpha}{I\bar{\nu}_\alpha}.
\ee
For the two {\it r}-modes mentioned previously, $b_\alpha$ is negative and therefore $x_\alpha$ is positive, and is interpreted as the fraction of accreted angular momentum that is fed back to the orbit by the resonant tidal interaction. It turns out that for both modes $x_\alpha$ is approximately equal to 0.5. The binary is more stable to mass transfer during the resonance locking regime, as the coefficient in front of a term driving an instability is reduced. Stability of the binary under mass transfer is governed by the evolution of the Roche lobe over-fill parameter $\Delta \equiv R_2 - R_L$, where $R_2$ is the donor radius and $R_L$ is the donor Roche lobe size. The equation governing the evolution of the over-fill parameter during resonance locking is given by

\bea
\frac{1}{2R_2}\frac{d\Delta}{dt} &=& \frac{1}{2}\left[(\zeta_2 - \zeta_{r\,L})\frac{\dot{M}_2}{M_2} - \frac{\dot{a}}{a}\right] \nn \\
&=& - \frac{\dot{J}_{\rm GW}}{J_{\rm orb}} - \frac{I}{\tau_S J_{\rm orb}}(\Omega - \omega_{\rm orb}) + \sigma\frac{\dot{M}_2}{M_2}, \label{Deltadot}
\eea
where $q$ is the mass ratio, $M_2$ is the donor mass and $\sigma$ is defined as

\be
\sigma = 1 + \frac{1}{2}(\zeta_2 - \zeta_{r_L}) - q - (1-x_\alpha)\sqrt{(1+q)R_h/a}.
\ee
The quantity $R_h$ is the radius of the orbit around the accretor that has the same specific angular momentum as the accreted mass, and can be conveniently approximated by the following expression (\cite{vr}) 

\be
\frac{R_h}{a} = 0.088 - 0.049\log q + 0.115\log^2 q + 0.020\log^3 q,
\ee
which is valid for the range $0.001 < q < 1$. For mass ratios ranging from 0.1 to 1, we have $R_h/a \sim 0.15$ in order of magnitude. Lastly we have

\be
\zeta_2 = \frac{d \log R_2}{d \log M_2}, 
\ee
\be
\zeta_{r_L} = \frac{d \log (R_L/a)}{d \log M_2},
\ee
with typical values $-0.6 < \zeta_2 < -0.3$ and $\zeta_{r_L} \simeq 1/3$. The binary will be unstable if mass transfer $\dot{M}_2 < 0$ increases the over-fill, since a larger overfill implies in turn a larger mass transfer and thus the process runs away, leading to the destruction of the binary. 

\begin{figure}
\hspace*{\fill}
\includegraphics[width=5.7cm,angle=270]{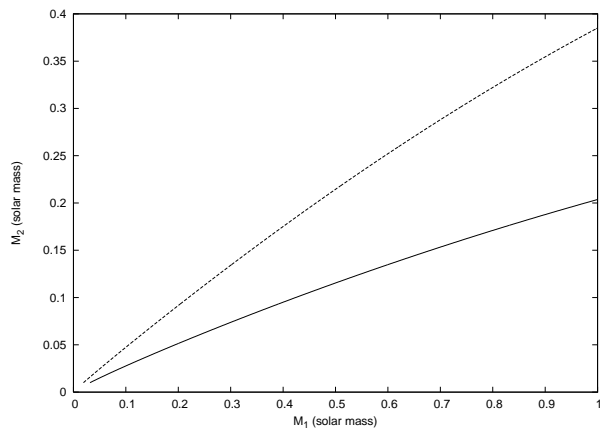}
\hspace*{\fill}
\caption{This figure shows the boundary of the regimes of guaranteed stability in two cases: (solid line) when no modes are driven, which is the criterion of Marsh, Nelemans and Steeghs (2004), and (dashed line) during resonance locking, which is Eq.(\ref{reslockstable}) with $x_\alpha = 0.5$. The binary is stable to mass transfer in the region of the graph below the line corresponding to the appropriate regime. Clearly resonance locking increases significantly the parameter space over which the binary can undergo stable equilibrium mass transfer.}\label{fig:stability}
\end{figure}

Let us now look for quasi-equilibrium solutions to (\ref{Deltadot}), i.e. solutions to $\dot{\Delta} = 0$.  The first term on the right-hand side of (\ref{Deltadot}) is positive and the second one is negative, since during resonance locking the (constant) spin frequency is larger than the orbital frequency. However in the limit of very weak dissipative tidal coupling ($\tau_S \sim 10^{15}$ yr), which we assume in this paper, the second term can be neglected and the criterion for existence a stable equilibrium solution is simply $\sigma > 0$, which translates to   

\be\label{reslockstable}
q < 1 + \frac{1}{2}(\zeta_2 - \zeta_{r_L}) - (1-x_\alpha)\sqrt{(1+q)R_h/a}.
\ee
Using Eggleton's zero-temperature mass-radius relation and his approximation for $\zeta_{r_L}$ as both quoted by \cite{marsh}, we can compare stability criterion (\ref{reslockstable}) to the stability criterion of \cite{marsh} [Eq.(31) of that paper]; the result is shown in Figure \ref{fig:stability}.

Now if $\sigma \leq 0$, we have $\dot{\Delta} > 0$ (still neglecting dissipative tidal coupling) and no stable equilibrium solution exists during resonance locking. In that case, the accretion rate will keep increasing until it becomes too strong for hydrodynamical spin stabilization to operate, and the accretor can potentially be spun up through the resonance without saturation of the mode amplitude. A precise modeling of the evolution of accretion rate during resonance locking is required to accurately quantify this effect, which we do not address in the present paper. 

It is also interesting to ask what the observational signature of the resonance locking regime would be, for example in the evolution of orbital frequency. Before resonance occurs the effect of tidal coupling due to modes on orbital angular momentum evolution is negligible. The equation governing the rate of change of orbital angular momentum in this regime is thus Eq.(\ref{JconsintroII}), with $x_\alpha$ taken to be zero. As resonance is approached, we show in section \ref{models} and figure \ref{fig:omega} that the accretor's spin gets stabilized over a timescale of about a year. Once the spin is stabilized the orbital angular momentum evolves according to (\ref{JconsintroII}) with $x_\alpha \simeq 0.5$ for the modes considered here. This implies that one should observe a sudden decrease $\delta \dot{\omega}_{\rm orb}$ in the time derivative of the orbital frequency $\dot{\omega}_{\rm orb}$ at the onset of resonance locking regime, given by

\bea
\delta \dot{\omega}_{\rm orb} &=& 3x_\alpha \frac{\omega_{\rm orb}}{J_{\rm orb}} \dot{J}_{\rm acc} \nn \\
&=& 3x_\alpha \frac{\omega_{\rm orb}}{J_{\rm orb}} \sqrt{M_1 R_h} \dot{M}_2 .\label{deltaomegaorb}
\eea
which is negative since $\dot{J}_{\rm acc} < 0$. Since $\dot{\omega}_{\rm orb}$ is negative because of mass transfer\footnote{Unless the binary is in accretion turn-on phase, where the mass transfer rate can be substantially below its equilibrium value, in which case it possible to have overall increasing orbital frequency.}, onset of resonance locking produces an increase in the magnitude of $\dot{\omega}_{\rm orb}$. Alternatively one may rewrite (\ref{deltaomegaorb}) in terms of orbital period derivative $\dot{P}_{\rm orb}$ as

\bea
\delta \dot{P}_{\rm orb} &=& - 3x_\alpha \frac{P_{\rm orb}}{J_{\rm orb}} \sqrt{M_1 R_h} \dot{M}_2 \nn \\
&\sim& 6\times10^{-5}\sqrt{\frac{1+q}{q^2}}\, \left(\frac{P_{\rm orb}}{10\, {\rm min}}\right) \left(\frac{M_1}{M_\odot}\right)^{-1} \nn \\
&& \times \left(\frac{|\dot{M}_2|}{10^{-5} M_\odot \, {\rm yr}^{-1}}\right)\,\, {\rm min}\,{\rm yr}^{-1},
\eea
where we have assumed $x_\alpha = 0.5$ and the order of magnitude $R_h \sim 0.15a$ to obtain the second line. Once the resonant mode saturates, the stabilization mechanism turns off and we may essentially reset $x_\alpha$ to zero in (\ref{JconsintroII}), leading to a rapid increase in $\dot{\omega}_{\rm orb}$, given by the negative of (\ref{deltaomegaorb}). Unfortunately this observational signature does not correspond to what is seen in RX J0806.3+1527 and V407 Vul, where the orbital frequency is observed to be increasing with time at a rate consistent with orbital dynamics being entirely dominated by gravitational wave radiation reaction. Resonance locking due to tidal excitation of {\it r}-modes tends to push the stars apart even further than mass transfer alone and thus provides no immediate help in identifying the exact nature of these two systems. However our hydrodynamical stabilization mechanism nevertheless plays a significant role in the orbital evolution of semi-detached ultra-compact binary white dwarfs. It must therefore be taken into account for identifying the region of parameter space where these systems are stable to mass transfer.

\subsection{Outline}

Our paper is structured as follows. We start in section \ref{sec:perttheory} by briefly reviewing the essential material from perturbation theory of rotating stars required for our analysis. We give a more detailed review in Appendix \ref{app:perttheory} for the reader interested in a streamlined introduction to the detailed, heavily mathematical formalism of \cite{Schenk}. In section \ref{sec:evolutionequations} we derive the evolution equation for the spin frequency of the accretor including tidal effects. This evolution equation is essentially a version of (\ref{totaltorque}) where $\dot{J}_{\rm tidal}$ and $J_2[\bmath{\xi},\bmath{\xi}]$ are given explicitly in terms of a mode expansion of the fluid displacement vector field. We also give the evolution equation for the orbital frequency, needed to justify approximations used later in section \ref{models}. In section \ref{models} we construct a simple dynamical system where we assume that a single mode is dynamically relevant for spin evolution. There we solve the equations of motion of this dynamical system both analytically and numerically, modeling the accretor as an incompressible, slowly rotating Maclaurin spheroid. In this section we also identify relevant candidate modes for stabilization and perform the estimate of the mechanism's lifetime.

\section{Elements of perturbation theory of rotating stars}\label{sec:perttheory}

In this paper we use perturbation theory of uniformly rotating stars as presented by \cite{Schenk}, which is very mathematical body of work. In this section we present the essential elements required for the analysis of this paper, and give a more detailed review for the interested reader in Appendix \ref{app:perttheory}. We use geometric units $G=c=1$ throughout this paper.

The perturbed stated of a uniformly rotating star can be entirely described by a Lagrangian fluid displacement vector field $\bmath{\xi}(\bmath{x},t)$, which tells how fluid elements are displaced from their background position $\bmath{x}$ in the frame corotating with the star. This fluid displacement is expanded in phase space as follows

\be\label{xiexpansion}
\left[\begin{array}{c} \bmath{\xi}(\bmath{x},t) \\ \bmath{\pi}(\bmath{x},t) \end{array} \right] = \sum_\alpha c_\alpha(t) \left[\begin{array}{c} \bmath{\xi}_\alpha(\bmath{x}) \\ \bmath{\pi}_\alpha(\bmath{x}) \end{array} \right] \, + {\rm c.c.} \,,
\ee
where $\bmath{\pi}$ is the momentum canonically conjugate to $\bmath{\xi}$. A given mode $\bmath{\xi}_\alpha$ with associated eigenfrequency $\omega_\alpha$ is solution to a specific eigenvalue equation [Eq.(\ref{eigenvalue})]. The mode amplitude coefficients $c_\alpha(t)$ obey the following evolution equation, which is deduced by substituting expansion (\ref{xiexpansion}) into Euler's equation,

\be\label{calphaevolution}
\dot{c}_\alpha + i\omega_\alpha c_\alpha = \frac{i}{b_\alpha}\int d^3x \, \rho \, \bmath{\xi}_\alpha^\ast(\bmath{x}) \cdot \bmath{a}_{\rm ext}(\bmath{x},t) ,
\ee
where $\rho$ is background mass density and $\bmath{a}_{\rm ext}$ is the acceleration field experienced by the fluid elements due to external perturbations, e.g. a companion object in a binary. The constant $b_\alpha$ is given by

\be
b_\alpha = 2\Omega i\int d^3x \, \rho \, \bmath{\xi}_\alpha^\ast \cdot \left(\hat{\bmath{\Omega}} \times \bmath{\xi}_\alpha\right) + 2\omega_\alpha \int d^3x \, \rho\, |\bmath{\xi}_\alpha|^2,
\ee
where $\Omega$ is the angular frequency of the star and $\hat{\bmath{\Omega}}$ is a unit vector pointing into the direction of the background star's angular momentum vector. The complete tidal response of a rotating star to a prescribed external perturbation is then obtained by solving (\ref{calphaevolution}) for all modes of the star and substituting the results in expansion (\ref{xiexpansion}).

\subsection{Adiabatic approximation for time-dependent spin frequency}\label{sec:adiab}

So far the formalism discussed here applies for constant background spin frequency $\bmath{\Omega}$. In accreting white dwarf binaries, the spin of the accretor will change in time so we need to discuss how to incorporate this in our analysis. A few complications arise in the case of time-varying spin. However if the spin of the star varies slowly in time compared to eigenfrequencies of dynamically relevant modes, we can still use normal modes that are solution of (\ref{eigenvalue}) in the following way. Consider the (infinite) sequence of vector spaces of solutions to (\ref{eigenvalue}) for some range of spin frequencies $(\Omega_{\rm min},\Omega_{\rm max})$, where $\Omega_{\rm min} > 0$ and $\Omega_{\rm max} < \Omega_{\rm break-up}$. Label the normal modes as $(\bmath{\xi}_\alpha(\bmath{x},\Omega),\omega_\alpha(\Omega))$, where $\Omega$ denotes which vector space the mode labeled with quantum numbers $\alpha$ belongs to\footnote{One does not necessarily need to consider this sequence of Hilbert spaces to analyze the case of time-dependent spin frequency. We refer the reader to Appendix A of \cite{btw} for a detailed discussion of perturbation theory of stars with generic time-dependent spin frequency.}. Assume now a displacement of the form $\bmath{\xi}(\bmath{x},t) = \exp\big[-i\int \omega_\alpha(\Omega) dt \big] \bmath{\xi}_\alpha(\bmath{x},\Omega)$. We then have
\be
\dot{\bmath{\xi}} = -i \omega_\alpha \bmath{\xi} + O(\dot{\Omega}  \bmath{\xi} / \Omega),
\ee
the second term being a measure of the change in the mode eigenfunction as the star is spun up or down\footnote{It is possible to make a continuous, one-to-one identification of modes between vector spaces of neighboring frequencies, as long as one is a finite distance away from $\Omega = 0$. The change in mode eigenfunctions can be meaningfully computed with the help of this identification.} . If the timescale of change of the spin frequency is much longer than the inverse normal mode frequency, the second term is negligible and we get
\be
\ddot{\bmath{\xi}} = - \omega_\alpha^2 \bmath{\xi} - i\dot{\Omega} \frac{\partial \omega_\alpha}{\partial \Omega} \bmath{\xi}.
\ee
Again if the timescale of change of the spin frequency is much longer than the inverse normal mode frequency, we can drop the second term in the right-hand side and we then see that our ansatz for the fluid displacement satisfies (\ref{eigenvalue}) with $\bmath{B} \rightarrow \bmath{B}(t)$, up to corrections of order $\dot{\Omega} / (\Omega \omega_\alpha)$, which we assume are much less than order unity\footnote{For compact white dwarf binaries with accretion rates of order $10^{-5} M_{\odot} \,{\rm yr}^{-1}$, spin frequency of order orbital frequency and mode frequency of order spin frequency, the ratio $\dot{\Omega}/\Omega \omega_\alpha$ is of order $10^{-8}-10^{-9}$. The adiabatic approximation is thus fully justified for such systems.}. In this adiabatic-type approximation, we can then expand a generic fluid displacement in phase space as follows
\be
\left[\begin{array}{c} \bmath{\xi}(\bmath{x},t) \\ \bmath{\pi}(\bmath{x},t) \end{array} \right] = \sum_\alpha c_\alpha(t) \left[\begin{array}{c} \bmath{\xi}_\alpha(\bmath{x},\Omega) \\ \bmath{\pi}_\alpha(\bmath{x},\Omega) \end{array} \right] \, + {\rm c.c.}
\ee
where the set $\{\bmath{\xi}_\alpha(\bmath{x},\Omega),\pi_\alpha(\bmath{x},\Omega)\}$ are the normal modes and their conjugate momenta obtained from perturbation theory of stars spinning at constant frequency. The time evolution of the mode amplitude coefficients is then given by (\ref{modeevolution}), again up to corrections of order $\dot{\Omega} / (\Omega \omega_\alpha)$, where now $\bmath{\xi}_\alpha$ and $b_\alpha$ depend slowly on time due to evolution of spin frequency. In the rest of this paper we will always omit the labels $\Omega$ on the modes and their frequencies. It will be understood that a given mode $(\bmath{\xi}_\alpha,\omega_\alpha)$ depends on spin frequency in the context of the adiabatic approximation detailed here.

\subsection{Mode amplitude evolution equation for tidal excitation in binary systems}

We now specialize to the case where the external acceleration vector $\bmath{a}_{\rm ext}$ is generated by the Newtonian gravitational field $\Phi_{\rm ext}$ of a companion object. We model the companion simply as a point particle of mass $M_2$ and assume that the orbital angular momentum is aligned with the spin axis of the perturbed star. The mode amplitude forcing term [the right-hand side of (\ref{modeevolution})] is then given by
\bea
f_\alpha &\equiv& \frac{i}{b_\alpha}\langle \bmath{\xi}_\alpha,\bmath{a}_{\rm ext} \rangle \nn \\
&=& -\frac{i}{b_\alpha}\int d^3x \, \rho \, \bmath{\xi}_\alpha^\ast \cdot \bmath{\nabla}\Phi_{\rm ext} \nn \\
&=& -\frac{i}{b_\alpha}\int d^3x \, \delta \rho_\alpha^\ast \, \Phi_{\rm ext}. \label{fAdeltarho}
\eea
where $\delta \rho_\alpha$ is the Eulerian density perturbation of mode $\alpha$. Since the background star is axisymmetric we may write the density perturbation due to mode $\alpha$ as
\be
\delta \rho_\alpha = g_\alpha(r,\theta) e^{im_\alpha\phi}.
\ee
In this paper we will use the convention $m_\alpha \geq 0$, which implies $\omega_\alpha$ may be positive or negative\footnote{The convention used in \cite{Schenk} is to choose $\omega_\alpha$ positive and thus differs from ours. If needed, see Appendix \ref{app:perttheory} for more details about mode pair notation.}.

Writing the orbital separation vector $\bmath{d}$ pointing from the accretor to the donor as $\bmath{d} = d(\cos \phi_{\rm orb},\sin\phi_{\rm orb},0)$. We then make use of the following expansion for the external Newtonian potential $\Phi_{\rm ext}$ (\cite{holai}) 
\be\label{Phiextexpansion}
\Phi_{\rm ext}(\bar{\bmath{x}},t) = -M_2 \sum_{l,m}W_{l m} \frac{r^{l}}{d^{l+1}}e^{-im \phi_{\rm orb}(t)} Y_{l m}(\theta, \varphi),
\ee
where $\bar{\bmath{x}}$ are inertial coordinates and where 
\bea
W_{l m} = (-)^{(l + m)/2} \left[\frac{4\pi}{2l + 1} (l+m)!(l-m)! \right]^{1/2}\nn \\
\times \left[2^{l}\left(\frac{l + m}{2}\right)!\left(\frac{l - m}{2}\right)!\right]^{-1},\label{Wlm}
\eea
the symbol $(-)^n$ being zero if $n$ is not an integer. Substituting expansion (\ref{Phiextexpansion}) into (\ref{fAdeltarho}) and using $\phi = \varphi - \int \Omega(t^\prime) dt^\prime$ then yields
\bea 
f_\alpha &=& \mathcal{F}_\alpha \exp\left[-im_\alpha\left(\phi_{\rm orb}(t) - \int^t \Omega(t^\prime) dt^\prime\right)\right] \nn \\
&\equiv& \mathcal{F}_\alpha e^{-im_\alpha u(t)}, \label{fAdef}
\eea
where $\mathcal{F}_\alpha$ is the following complex number
\bea
\mathcal{F}_\alpha = \frac{iM_2}{b_\alpha} \sum_{l = m_\alpha}^\infty \int d^3x \, g_\alpha^\ast(r,\theta) W_{l m_\alpha} \frac{r^{l}}{d^{l+1}} \nn \\
\times Y_{l m_\alpha}(\theta,\phi)e^{-im_\alpha\phi}. \label{FsubA}
\eea
The mode amplitude then satisfies 
\be\label{modeampEOMII}
\dot{c}_\alpha + i\omega_\alpha c_\alpha = \mathcal{F}_\alpha e^{-im_\alpha u(t)}.
\ee
If the force term oscillates at a frequency far from the mode's normal frequency, we may write the following approximate solution to (\ref{modeampEOMII})
\be\label{cAapproxI}
c_\alpha = \frac{i\mathcal{F}_\alpha e^{-im_\alpha u(t)}}{m_\alpha\dot{u} - \omega_\alpha}.
\ee
Equation (\ref{cAapproxI}) is only valid if the mode has not encountered a resonance at an earlier time, in which case (\ref{cAapproxI}) needs to be supplemented by a homogeneous solution to (\ref{modeampEOMII}) obtained by matching to the solution to (\ref{modeampEOMII}) during resonance (\cite{me}). 

\section{Evolution of spin and orbital frequencies}\label{sec:evolutionequations}

In this section we give evolution equations for the spin frequency of the perturbed star and the orbital frequency of a binary undergoing conservative mass transfer. These quantities are essential in solving the mode amplitude equation of motion (\ref{modeampEOMII}) since they determine the phase $u(t)$ of the forcing term. 

\subsection{Evolution of spin frequency}\label{sec:spinevol}

We obtain the evolution equation for the spin frequency from the fundamental equation

\be\label{Jstarcons}
\frac{dJ_{\rm star}}{dt} = \mathcal{T}_{\rm ext},
\ee
where $\mathcal{T}_{\rm ext}$ is the external torque acting on the star. In Appendix \ref{app:spinevol} we show explicitly that the total angular momentum of the perturbed star is given by

\be
J_{\rm star} = I_0\Omega_0 + \left[1 + \frac{d\ln I_0}{d\ln \Omega_0}\right]I_0\delta \Omega + \int d^3x \, \rho \, \hat{\Omega}\cdot(\bmath{\xi} \times \bmath{\pi}),
\ee
where $I_0$ is the unperturbed star's moment of inertia, $\Omega_0$ is the unperturbed star's spin frequency, and $\delta \Omega$ is the perturbation in the star's spin frequency, so that the full spin frequency is $\Omega = \Omega_0 + \delta \Omega$. Assuming that the change in the star's moment of inertia due to changes in spin and mass (because of accretion) can be neglected, (\ref{Jstarcons}) then gives

\be\label{spinfrequencyevol}
\frac{d\Omega}{dt} = \frac{1}{I}\left[\mathcal{T}_{\rm ext} - \frac{d}{dt} \int d^3x \, \rho \, \hat{\Omega}\cdot(\bmath{\xi} \times \bmath{\pi})\right],
\ee 
where $I$ now stands for the moment of inertia of a non-spinning star, as we neglect the effect of spin on $I$. Equation (\ref{spinfrequencyevol}) is the desired evolution equation for the spin frequency. In Appendix \ref{app:spinevol}, we derive a form of the spin evolution equation in terms of mode amplitude coefficients, which is given by Eq.(\ref{OmegadotwithIeff_app}) below.

At this point of our analysis it is worth discussing the compatibility of Eq.(\ref{dotOmegafinal}) with the theorem of \cite{GN}, who show that the secular change in specific angular momentum of a fluid element following a suitably defined "background trajectory" vanishes if the star is subject to the gravitational perturbation of a companion object rotating around the star. We want to point out here that this theorem does not necessarily imply that the spin of the star cannot be altered, in the sense discussed in section \ref{sec:Jordanchains}, by tidal interactions. This important point can be illustrated simply as follows. Label by $\bmath{x}_0$ the location at $t=0$ of a fluid element of the unperturbed star. We denote the trajectory this fluid element would follow if the star were to remain unperturbed at all times by $\bmath{x}(\bmath{x}_0,t)$. Assume now that at $t=0$ a perturbation is turned on and that the exact state of the star is described by fluid elements following the trajectory $\bmath{y}(\bmath{x}_0,t)$. In the language of perturbation theory described in this paper, the fluid displacement vector is defined as
\be
\bmath{y}(\bmath{x}_0,t) = \bmath{x}(\bmath{x}_0,t) + \bmath{\xi}(\bmath{x}_0,t).
\ee
Note however that the trajectory $\bmath{x}(\bmath{x}_0,t)$ does not describe the background state defined in \cite{GN}. They instead decompose the exact state as follows
\be
\bmath{y}(\bmath{x}_0,t) = \bmath{x}^\ast(\bmath{x}_0,t) + \bmath{\xi}^\ast(\bmath{x}_0,t),
\ee
where the time average of the perturbation $\bmath{\xi}^\ast$ about the background trajectory $\bmath{x}^\ast$ is required to vanish. Now in general, the time average of the perturbation $\bmath{\xi}$ about the unperturbed trajectory $\bmath{x}$ does not vanish, one specific example being when a mode of the star is resonantly excited. Hence what Goldreich and Nicholson call the background state in general differs from the unperturbed state. To linear order in $\bmath{\xi}$, we have
\be
\bmath{x}^\ast(\bmath{x}_0,t) = \bmath{x}(\bmath{x}_0,t) + \langle \bmath{\xi}(\bmath{x}_0,t) \rangle + O(\bmath{\xi}^2),
\ee
where here the brackets $\langle \, \rangle$ denote a suitably defined time averaging procedure about the unperturbed state (\cite{GN}). The point is that the quantity $\langle \bmath{\xi} \rangle$ contains in general differential rotation modes and therefore the background state described by fluid worldlines $\bmath{x}^\ast$ does not generically describe a fluid in uniform (rigid) rotation. One therefore may not interpret the theorem of \cite{GN} as implying that tidal interactions cannot change the spin of a star, in the sense defined in section \ref{sec:Jordanchains}. Still the theorem of \cite{GN} may in principle be converted into an evolution equation for the secular change of the spin frequency if one decomposes the background state onto the basis of Jordan chain modes describing arbitrary differential rotation and then isolates the rate of change of the coefficient $c_1$ of the uniform rotation Jordan chain mode. 

Now in the special case where $\langle \bmath{\xi} \rangle$ vanishes, then the background state of Goldreich and Nicholson does correspond to the unperturbed state (which here is a uniformly rotating star). In that case their theorem does indeed imply that the secular change in the spin frequency of the star vanishes, in the context of a star perturbed by an orbiting companion object. We can easily show explicitly that this result is compatible with the equation of motion (\ref{dotOmegafinal}) we derived rigorously using perturbation theory of uniformly rotating stars. To carry out this proof one first makes the ansatz $\dot{\Omega} = 0$ to solve the mode amplitude evolution equation (\ref{modeampEOMII}). One then finds that $c_A(t) = |c_A|e^{-im_Au(t)}$, where $|c_A|$ is constant. Clearly the time average of $c_A(t)$ over a timescale long compared to $1/\dot{u}$ is zero and therefore $\langle \bmath{\xi} \rangle = 0$. Using this form of the mode amplitude coefficients, the form (\ref{tidaltorquefinal}) for the total external torque applied on the star and averaging Eq.(\ref{dotOmegafinal}) over a timescale long compared to $1/\dot{u}$, one finds immediately that the time averaged change in the spin frequency vanishes, consistent with the ansatz used to solve the mode amplitude evolution equation. Thus a constant spin frequency is the solution to the time averaged spin evolution equation, consistent with the theorem of Goldreich and Nicholson. However in more general situations where the frequency $\dot{u}$ of the driving term in the mode amplitude evolution equation (\ref{modeampEOMII}) varies in time, leading possibly to resonant excitations of normal modes, one obtains the time evolution of the spin frequency of the star by solving simultaneously (\ref{modeampEOMII}) and (\ref{dotOmegafinal}), as long as the adiabatic approximation of section \ref{sec:adiab} is valid.

\subsection{Orbital frequency evolution}\label{sec:dotomegaorb}

We next give the evolution equation for the orbital frequency $\omega_{\rm orb}$. Below we assume a Newtonian quasi-circular orbit, i.e. $\omega_{\rm orb}^2 = (M_1+ M_2)/d^3 \equiv M/d^3$, with $M$ being constant. We follow \cite{marsh} closely. We assume the orbital angular momentum $J_{\rm orb}$ evolves due to three effects, namely emission of gravitational waves, mass transfer and tidal coupling. We may then write
\be
\dot{J}_{\rm orb} = \dot{J}_{\rm GW} + \dot{J}_{\rm acc} + \dot{J}_{\rm tidal},
\ee
where 
\be
\dot{J}_{\rm GW} = - \frac{32}{5d^4}\mu M^2 J_{\rm orb}.
\ee
In Appendix \ref{tidalcouplingtorque}, we review the computation of $\dot{J}_{\rm tidal}$ in Newtonian gravity and show that it can be written to leading order in the fluid perturbation as a sum over modes as follows
\be
\dot{J}_{\rm tidal} = -\sum_{\alpha} m_\alpha b_\alpha \frac{d}{dt} |c_\alpha|^2 + \frac{1}{\tau_S} I (\Omega - \omega_{\rm orb}),
\ee
the second term in the right-hand side modeling dissipative tidal coupling. As noted in Appendix K of \cite{Schenk}, each individual term of the above sum does {\it not} correspond to the angular momentum deposited in mode $\alpha$; the term $J_2[\bmath{\zeta},\bmath{\zeta}]$ [cf. Eq.(\ref{Jstarexpanded})] giving the angular momentum in the star carried by the perturbation contains in general cross-terms between different modes. For a quasi-circular orbit and conservative mass transfer we have
\be
\dot{J}_{\rm orb} = \left[-\frac{\dot{\omega}_{\rm orb}}{3\omega_{\rm orb}} + (1-q)\frac{\dot{M}_2}{M_2}\right]J_{\rm orb},
\ee
where $q$ is the mass ratio $M_2 / M_1$. We then obtain the following evolution equation for orbital frequency
\bea
\dot{\omega}_{\rm orb} &=& -3\omega_{\rm orb}\left[-\frac{32}{5d^4}\mu M^2 + \frac{\dot{J}_{\rm acc}}{J_{\rm orb}} - (1-q)\frac{\dot{M}_2}{M_2} \right. \nn \\ 
&& \left. - \frac{1}{J_{\rm orb}}\sum_{\alpha} m_\alpha b_\alpha \frac{d}{dt} |c_\alpha|^2 + \frac{I(\Omega - \omega_{\rm orb})}{J_{\rm orb}\tau_S} \right]. \label{dotomegaorb}
\eea
Comparing (\ref{dotOmegafinal}) and (\ref{dotomegaorb}) and assuming that the binary is in a regime where $\Omega \sim \omega_{\rm orb}$, i.e. close to synchronization, we then see that roughly, $\dot{\omega}_{\rm orb} \sim (R/d)^2\dot{\Omega}$. Thus for the purpose of solving (\ref{modeampEOMII}), we will assume as a first approximation that $\omega_{\rm orb}$ is constant, since $\ddot{u}(t)$ is dominated by spin frequency time evolution. We shall leave including back-reaction on orbital parameters when solving for the mode amplitude for a future publication.

\section{A single generalized Rossby mode as synchronizing mechanism}\label{models}

In this section we develop and analyze a simple dynamical system that attempts to extract the dominant features of the synchronization mechanism we propose, namely the stabilization of the accreting star's spin frequency through resonant tidal excitation of normal modes. The essential features of the physical system we model in this section may be summarized as follows. To begin with consider a compact, detached binary white dwarf system. The binary is compact enough that it coalesces due to gravitational wave emission. The system eventually reaches a point where the larger, less massive white dwarf fills its Roche lobe and mass transfer begins. The accreting white dwarf spins up as it accumulates material from its companion and eventually reaches a spin frequency at which one of its normal modes is resonantly driven. The amplitude of the mode then grows rapidly and may affect significantly the evolution of the spin frequency, depending on the size of mode back-reaction [cf. Eq.(\ref{dotOmegafinal})]. 

Our aim is to solve the coupled dynamical equations (\ref{modeampEOMII}) and (\ref{dotOmegafinal}) for mode amplitude and spin frequency for a single resonant mode and use this solution to catalog which modes are suitable candidates for stabilizing the spin of the accretor. In this paper we will focus solely on the excitation of generalized Rossby modes (\cite{bryan,lindblom}), since their normal frequencies scale as the star's spin frequency, and may then resonate when the star's spin frequency is of order orbital frequency. Other modes like {\it f} and {\it p} modes resonate only at spin frequencies much larger than synchronized spin frequency. It is then likely that dissipative tidal coupling prevents the accretor to reach a large enough spin for such modes to stabilize it. For stars with buoyancy, {\it g}-modes are also potential candidates for spin stabilization at spin frequency of order orbital frequency, but we will not investigate them in this paper. In Appendix \ref{modeidentities} we review the properties of generalized {\it r}-modes needed for our purposes. For a modern, in-depth discussion of these modes, the interested reader is invited to consult \cite{lindblom}.
 
\subsection{Main assumptions and evolution equations}
 
The dynamical system we study in this section is of course a crude approximation to the true dynamics of the binary. The main assumptions we make when solving (\ref{modeampEOMII}) and (\ref{dotOmegafinal}) are the following. 

\begin{enumerate}

\item First we assume, as mentioned at the end of section \ref{sec:dotomegaorb}, that orbital frequency is constant for the purpose of solving the mode amplitude evolution equation (\ref{modeampEOMII}).

\item We assume the accretor spins slowly enough so that (a) we may use appropriate approximate formulae for the {\it r}-mode eigenfunctions and eigenfrequencies and (b) we may consider its moment of inertia as constant.

\item We assume a constant torque on the star due to accretion of material from the companion.

\item We assume that the Rossby modes are driven solely by the gravitational tidal field of the companion, i.e. we neglect the external forcing due to impact of infalling material on the surface star when computing the driving term in Eq.(\ref{modeampEOMII}).

\item Lastly we assume that the evolution of the spin frequency described by Eq.(\ref{spinfrequencyevol}) is dominated by a single {\it r}-mode.

\end{enumerate}

A given {\it r}-mode may be a suitable candidate for tidal synchronization only if it satisfies the following four criteria: (i) its azimuthal quantum number $m$ must be non-zero, otherwise the tidal driving term does oscillate in time [cf. Eq.(\ref{modeampEOMII})] and resonance is not possible; (ii) its quantum numbers must satisfy 
\be
l+m =  {\rm even}, 
\ee
otherwise the mode does not couple to the Newtonian tidal field [cf. Eqs.(\ref{Wlm}),(\ref{FsubA}) and (\ref{deltarhoslowrot})] \footnote{Note that this condition excludes the so-called classical {\it r}-modes since they have $l = m+1$. This restriction on $l+m$ is lifted if we allow misalignment of the spin and the orbital angular momentum. However since we expect most of the white dwarf spin to come from accretion, the amount of misalignment should be small and tidal coupling of {\it r}-modes with $l+m$ odd should be suppressed by some power of this small misalignment angle.}; (iii) for $m > 0$ the mode's dimensionless frequency $w = \omega/2\Omega$ must satisfy the condition
\be
\lambda \equiv m + 2w > 0
\ee
for resonance to be at all possible, as can be seen from (\ref{omegaresdef}), and (iv) the mode's back-reaction parameter $\bar{\nu}_\alpha$, encountered in Eq.(\ref{freqevolintro}) and defined precisely by Eqs. (\ref{taudef}) and (\ref{barnudef}) below, must be positive and large enough (how large will be determined later in subsection \ref{approxanalytic}) for the mode to be a candidate for tidal synchronization. As shown in table \ref{table:numericalvalues} in Appendix \ref{integrals}, there are five candidate modes, with $l \leq 6$, satisfying criteria (i)-(iii). Each candidate mode is marked by an asterisk in that table. We shall see below that modes will $l \geq 7$ are not expected to satisfy criterion (iv) for generic binaries, which is why they are excluded from table \ref{table:numericalvalues}.

As mentioned above our model dynamical system consists of two coupled differential equations, one for the (complex) mode amplitude $c_\alpha$ and one for the spin frequency of the star $\Omega$. For a single mode, spin frequency evolution equation (\ref{spinfrequencyevol}) reduces to
\bea
\dot{\Omega} &=& \frac{1}{I} \left[\mathcal{T}_{\rm ext} - 2\left\langle \hat{\bmath{\Omega}}\times\bmath{\xi}_\alpha,-i\omega_\alpha\bmath{\xi}_\alpha + \bmath{\Omega} \times \bmath{\xi}_\alpha \right\rangle \frac{d}{dt}|c_\alpha|^2\right] \nn \\
&\equiv& \frac{1}{I} \left[\mathcal{T}_{\rm ext} - \kappa_\alpha\frac{d}{dt}|c_\alpha|^2\right]\label{taudef}
\eea
Conservation of angular momentum yields
\bea
\mathcal{T}_{\rm ext} &=& -\dot{J}_{\rm tidal} - \dot{J}_{\rm acc} \nn \\ 
&=& m_\alpha b_\alpha \frac{d}{dt}|c_\alpha|^2 - \dot{J}_{\rm acc},
\eea
where $\dot{J}_{\rm tidal}$ and $\dot{J}_{\rm acc}$ are the rates of change of orbital angular momentum due tidal interactions and accretion respectively. We then obtain the following system of equations
\be
\dot{\Omega} = \dot{\Omega}_{\rm acc} - \bar{\nu}_\alpha \frac{d}{dt}|c_\alpha|^2,  \label{spinevoleq}
\ee
\be
\dot{c}_\alpha + i\omega_\alpha c_\alpha = \mathcal{F}_\alpha e^{-im_\alpha u}, \label{modeevoleq}
\ee
where the quantities $\dot{\Omega}_{\rm acc}$ and $\bar{\nu}_\alpha$ are given by 
\be
\dot{\Omega}_{\rm acc} = - \frac{1}{I}\dot{J}_{\rm acc}, 
\ee
\be
\bar{\nu}_\alpha = \frac{1}{I}\big(\kappa_\alpha - m_\alpha b_\alpha\big), \label{barnudef}
\ee
where $\dot{\Omega}_{\rm acc}$ is assumed constant. For the analysis of system (\ref{spinevoleq})-(\ref{modeevoleq}) performed in the next subsection, the only information we need about the parameters $\mathcal{F}_\alpha$ and $\bar{\nu}_\alpha$ is that they are of the form $({\rm constant}) \times \Omega$. Explicit expressions for $\mathcal{F}_\alpha$ and $\bar{\nu}_\alpha$ for generalized {\it r}-modes of slowly rotating stars are provided in Appendix \ref{integrals}. 

\subsection{Approximate analytic solution}\label{approxanalytic}

Let us start our analytic investigation of system (\ref{spinevoleq})-(\ref{modeevoleq}) by defining a few quantities as follows
\be
\gamma_\alpha \equiv \frac{c_\alpha}{\mathcal{F}_\alpha}e^{im_\alpha u},
\ee
\be
\tilde{\nu}_\alpha \equiv \frac{1}{\Omega^3}\bar{\nu}_\alpha |\mathcal{F}_\alpha|^2, \label{tildenudef}
\ee
\be
\lambda_\alpha \equiv m_\alpha + 2w_\alpha ,
\ee
\be
\Omega_\alpha^{\rm res} \equiv \frac{m_\alpha}{\lambda_\alpha}\omega_{\rm orb}.
\ee 
Definition (\ref{tildenudef}) is such that parameter $\tilde{\nu}_\alpha$ is dimensionless and constant. We can then rewrite the system (\ref{spinevoleq})-(\ref{modeevoleq}) as follows
\bea
\dot{\Omega} = \dot{\Omega}_{\rm acc} - \left[\tilde{\nu}_\alpha \Omega^3 \frac{d}{dt}|\gamma_\alpha|^2 \right]\times\big[1 + O(\dot{\Omega}^{1/2}/\Omega)\big], \label{spinevoleqprime}\\
\dot{\gamma}_\alpha + \Big[i\lambda_{\alpha}\big(\Omega - \Omega_\alpha^{\rm res}\big) \gamma_\alpha\Big]\times\big[1 + O(\dot{\Omega}/\Omega^2)\big] = 1. \label{modeevoleqprime}
\eea
The scaling of the error terms in (\ref{spinevoleqprime}) comes from assuming that the timescale for evolution of $|\gamma_\alpha|^2$ is at most the "no back-reaction" resonance timescale $t_{\rm res} \sim \dot{\Omega}^{-1/2}$. The main advantage of this formulation of the evolution equations is that we have gotten rid of the term $e^{-im_\alpha u}$ in the mode amplitude equation of motion. However the presence of the $\Omega^3$ factor in the evolution equation (\ref{spinevoleqprime}) for the spin still makes the equation very complicated to solve. But since we expect back-reaction of the modes on the spin to be dynamically relevant mostly in the vicinity of the resonance, we simplify (\ref{spinevoleqprime}) by replacing $\Omega$ by $\Omega^{\rm res}_\alpha$ in the second term of the right-hand side. We can estimate the errors introduced by this approximation as follows. Away from resonance the solution for the rescaled mode amplitude $\gamma_\alpha$ is given by
\bea
|\gamma_\alpha|^2 &=& \frac{1}{\lambda_\alpha^2(\Omega - \Omega_\alpha^{\rm res})^2} \left[1 + O\left(\frac{\dot{\Omega}}{\Omega\lambda_\alpha(\Omega - \Omega_\alpha^{\rm res})}\right) \right. \nn \\
&& \left.  + O\left(\frac{\dot{\Omega}}{\lambda_\alpha^2(\Omega - \Omega_\alpha^{\rm res})^2}\right)\right], \label{gamma2approx}
\eea
yielding simply
\be\label{Omegaaway}
\dot{\Omega} = \dot{\Omega}_{\rm acc}[1 + O(\tilde{\nu}_\alpha)]
\ee
in that regime. Numerically the $O(\tilde{\nu}_\alpha)$ corrections lie within the range $10^{-5} - 10^{-7}$ for modes of interest, independently of whether or not $\Omega$ is replaced by $\Omega_\alpha^{\rm res}$ in the right-hand side of (\ref{spinevoleqprime}). Thus replacing $\Omega$ by $\Omega^{\rm res}_\alpha$ in the second term of the right-hand side of (\ref{spinevoleqprime}) introduces negligible corrections away from resonance. Now approximations (\ref{gamma2approx}) and (\ref{Omegaaway}) fail when resonance regime is reached, which is when the dominant error terms in (\ref{gamma2approx}) become of order unity, i.e.
\be
\Omega - \Omega_\alpha^{\rm res} \sim \dot{\Omega}^{1/2},
\ee
or alternatively when
\be
\Omega \sim \Omega_\alpha^{\rm res}\left[1 + O\left(\frac{\dot{\Omega}^{1/2}}{\Omega_\alpha^{\rm res}}\right)\right].
\ee
For accreting binary white dwarfs these correction terms are typically of order $10^{-4}$. Thus approximating $\Omega$ by $\Omega_\alpha^{\rm res}$ in the right-hand side of (\ref{spinevoleqprime}) introduces negligible corrections also in the resonance regime. Defining
\be\label{nudef}
\nu_\alpha = \tilde{\nu}_\alpha \left(\Omega_\alpha^{\rm res}\right)^3,
\ee
the system of equations we solve here can be written in its final form as
\be
\dot{\Omega} = \dot{\Omega}_{\rm acc} - \nu_\alpha \frac{d}{dt}|\gamma_\alpha|^2 , \label{spinevoleqII}
\ee
\be
\dot{\gamma}_\alpha + i\lambda_{\alpha}\big[\Omega - \Omega_\alpha^{\rm res}\big] \gamma_\alpha = 1. \label{modeevoleqII}
\ee
The initial conditions we impose to integrate system (\ref{spinevoleqII})-(\ref{modeevoleqII}) are the following. We assume accretion begins at $t=0$ and that the accretor is not spinning prior to accreting phase. This implies of course $\Omega(0) = 0$. For the mode amplitude we assume that it starts at its equilibrium tide value, namely $\gamma_\alpha(0) = i/\lambda_{\alpha}\Omega_\alpha^{\rm res}$, obtained by setting $\dot{\gamma}_\alpha = 0$ at $t=0$ in (\ref{modeevoleqII}). We are now ready to solve (\ref{spinevoleqII})-(\ref{modeevoleqII}) analytically. Writing mode amplitude $\gamma_\alpha$ as
\be
\gamma_\alpha = \gamma_\alpha^R + i \gamma_\alpha^I,
\ee
with $\gamma_\alpha^{R,I}$ real, we may rewrite (\ref{spinevoleqII}) using (\ref{modeevoleqII}) as follows
\be\label{spinevoleqIII}
\dot{\Omega} = \dot{\Omega}_{\rm acc} - 2\nu_\alpha \gamma_\alpha^R.
\ee
Taking a time derivative of (\ref{spinevoleqIII}) and using (\ref{modeevoleqII}) once more yields
\be
\ddot{\Omega} + 2\nu_\alpha = -h_\alpha  \big(\Omega - \Omega_\alpha^{\rm res}\big) \label{ddotOmega},
\ee
where the quantity $h_\alpha$ is defined as
\be
h_\alpha = 2\nu_\alpha\lambda_\alpha \gamma_\alpha^I.
\ee
Then the imaginary part of (\ref{modeevoleqII}) may be rewritten as follows 
\be\label{hdot}
\dot{h}_\alpha = -\lambda_\alpha^2(\Omega - \Omega_\alpha^{\rm res})(\dot{\Omega}_{\rm acc} - \dot{\Omega}).
\ee
So far all that has been done is rewriting system (\ref{spinevoleq})-(\ref{modeevoleq}) in a slightly different, but very useful form. We are now ready to identify the crucial approximation needed to solve (\ref{ddotOmega}) and (\ref{hdot}) simultaneously. The key lies in realizing that for a large enough value of $\nu_\alpha$, which will be determined below, we may discard the $\ddot{\Omega}$ term on the left-hand side of (\ref{ddotOmega}). This approximation essentially yields some average spin frequency since we are dropping a term that generates oscillations in $\Omega$ at frequency $\sim \sqrt{h_\alpha}$. This average frequency solution is obtained directly from Eq.(\ref{ddotOmega}) and is equal to
\be\label{omegalargenu}
\Omega = \Omega_\alpha^{\rm res} - \frac{2\nu_\alpha}{h_\alpha}
\ee
Substituting (\ref{omegalargenu}) into (\ref{hdot}) then gives
\be\label{hdotII}
\dot{h}_\alpha = \frac{2\lambda_\alpha^2\nu_\alpha \dot{\Omega}_{\rm acc}h_\alpha^2}{h_\alpha^3 + 4\lambda_\alpha^2\nu_\alpha^2},
\ee
which is then easily integrated. The result is the following cubic equation for $h_\alpha$
\[
h_\alpha^3 - \left[4\lambda_\alpha^2\nu_\alpha (\dot{\Omega}_{\rm acc}t - \Omega_\alpha^{\rm res} ) + \frac{4\nu_\alpha^2}{(\Omega_\alpha^{\rm res})^2} \right] h_\alpha - 8\lambda_\alpha^2\nu_\alpha^2 
\]
\be
\equiv h_\alpha^3 - a_1(t)h_\alpha - a_0 = 0. \label{hcubic}
\ee
The average frequency is then given by (\ref{omegalargenu}), choosing the unique root of (\ref{hcubic}) that satisfies the initial condition $h_\alpha(0) = 2\nu_\alpha / \Omega_\alpha^{\rm res}$. The explicit form of the complete solution is
\bea
\Omega(t) &=& \Omega_{\rm res} - 2\nu_\alpha\left\{\theta(t_c - t)\left[\frac{a_0}{2} + \sqrt{\frac{a_0^2}{4} - \frac{a_1(t)^3}{27}}\right]^{1/3} \right. \nn \\
&& + 2\theta(t-t_c)\sqrt{\frac{a_1(t)}{3}}\cos\left[\frac{1}{3}\arccos\left(\frac{3^{3/2}a_0}{2a_1(t)^{3/2}}\right)\right] \nn \\ 
&& \left. + \theta(t_c - t)\left[\frac{a_0}{2} - \sqrt{\frac{a_0^2}{4} - \frac{a_1(t)^3}{27}}\right]^{1/3} \right\}^{-1},\nn \\\label{fullsol}
\eea
where the critical time $t_c$ is defined such that
\be
a_1(t_c) = \left(\frac{27a_0^2}{4}\right)^{1/3}.
\ee
Clearly solution (\ref{fullsol}) is not particularly revealing. In the large $t$ limit however solution (\ref{fullsol}) assumes the following very simple form
\be\label{Omegalarget}
\Omega(t) = \Omega_\alpha^{\rm res}  - \left[\frac{\nu_\alpha}{\lambda_\alpha^2\dot{\Omega}_{\rm acc}(t-t_0)}\right]^{1/2} \,\,\,\,\, {\rm for }\,\,\,\,\, t \gg t_0,
\ee
where
\be
t_0 = \frac{\Omega_\alpha^{\rm res}}{\dot{\Omega}_{\rm acc}} - \frac{\nu_\alpha}{\lambda_\alpha^2(\Omega_\alpha^{\rm res})^2\dot{\Omega}_{\rm acc}} \simeq \frac{\Omega_\alpha^{\rm res}}{\dot{\Omega}_{\rm acc}}.
\ee
The constant $t_0$ is essentially the time it would take to spin the star up to the resonant frequency  in the absence of mode back-reaction.

It should be clear that approximation (\ref{omegalargenu}), leading to the large $t$ solution (\ref{Omegalarget}), must necessarily fail when back-reaction parameter $\nu_\alpha$ is small, since in that case the star spins up through the resonance. An argument that shows this goes as follows. Assume (\ref{omegalargenu}) and (\ref{hdotII}) to be valid at $t=0$. Initially we have $h_\alpha > 0$ and hence $\dot{h}_\alpha > 0$. Equation (\ref{hdotII}) then implies $ h_\alpha > h_\alpha(0)  > 0$ for all time, which means that $\Omega < \Omega_\alpha^{\rm res}$ for all time, which cannot be true for $\nu_\alpha$ less than some critical value $\nu_{\rm crit}$, since in the limit $\nu_\alpha \rightarrow 0$, the solution for the spin approaches $\Omega(t) \rightarrow \dot{\Omega}_{\rm acc} t$ continuously. Approximation (\ref{omegalargenu}) can therefore work well for all time only for modes that will stabilize the spin frequency, i.e. modes with $\nu_\alpha > \nu_{\rm crit}$. The value of $\nu_{\rm crit}$ may then be estimated by finding when approximation (\ref{omegalargenu}) fails, which is when $\ddot{\Omega} \sim 2\nu_\alpha$. Using (\ref{omegalargenu}) and (\ref{hdotII}), we have
\be\label{basiccrit}
\frac{\ddot{\Omega}}{{2\nu_\alpha}} = -3(2\lambda_\alpha^2\nu_\alpha  \dot{\Omega}_{\rm acc})^2\frac{h_\alpha^4}{(h_\alpha^3 + 4\lambda_\alpha^2\nu_\alpha^2)^3}
\ee
This expression has a maximum at $h_\alpha^3 = 16\lambda_\alpha^2\nu_\alpha^2 / 5$. Putting this value for $h_\alpha$ in (\ref{basiccrit}) and requiring $\ddot{\Omega} \sim 2\nu_\alpha$ then yields the following value for the critical value $\nu_{\rm crit}$ of the back-reaction parameter
\be\label{nucrit}
\nu_{\rm crit} = 0.24 \sqrt{\lambda_\alpha} \dot{\Omega}_{\rm acc}^{3/2}.
\ee
Along with the late time solution (\ref{Omegalarget}) for spin frequency, equation (\ref{nucrit}) is one of our key results. Again only modes with $\nu_\alpha \gg \nu_{\rm crit}$ may effectively stabilize the spin of the accretor through tidal resonance. Modes with back-reaction parameters $\nu_\alpha \la \nu_{\rm crit}$ do not have a strong enough stabilizing effect on the spin to prevent the accretor from being spun up through the resonance. For the candidate modes shown in table \ref{table:numericalvalues}, $\sqrt{\lambda_\alpha} \la 1.7$ so a more conservative form of (\ref{nucrit}) that we may apply to all candidate modes is $\nu_{\rm crit} = 0.4\, \dot{\Omega}_{\rm acc}^{3/2}$. Another useful version of bound (\ref{nucrit}) that must be satisfied by spin stabilizing modes is the following
\be
\frac{\bar{\nu}_\alpha}{\Omega_\alpha^{\rm res}} \gg  0.24 \sqrt{\lambda_\alpha} \left|\frac{\mathcal{F}_\alpha}{\Omega_\alpha^{\rm res}}\right|^{-2} \left(\frac{\dot{\Omega}_{\rm acc}^{1/2}}{\Omega_\alpha^{\rm res}}\right)^3. \label{barnubound}
\ee
The latter version of the critical bound highlights better its dependence on tidal coupling strength, as $\bar{\nu}_\alpha / \Omega_\alpha^{\rm res}$ is independent of binary parameters while the back-reaction parameter $\nu_\alpha$, to which bound (\ref{nucrit}) refers to, depends on those parameters.

\subsection{Numerical results}

\begin{figure}
\hspace*{\fill}
\includegraphics[width=5.7cm,angle=270]{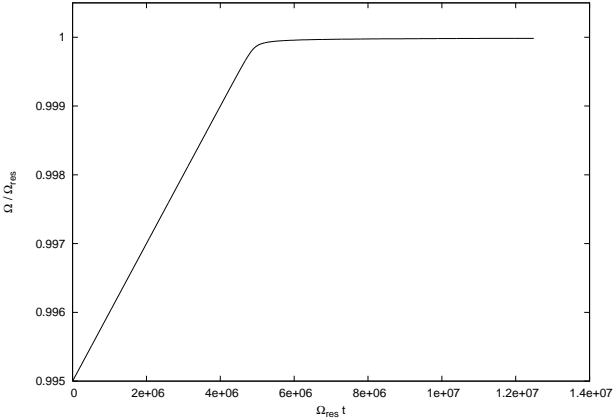}
\hspace*{\fill}
\caption{This figure shows the time evolution of accretor's spin frequency obtained by integrating system (\ref{spinevoleq})-(\ref{modeevoleq}) numerically. The following model parameters were used for moment of inertia of the accretor, the tidal ratio, the mass ratio and accretion-induced spin-up rate respectively:  $I = 0.4 MR^2$, $R/d = 0.2$, $q=0.25$ and $\dot{\Omega}_{\rm acc} = 10^{-9}$.}\label{fig:omega}
\end{figure}

Here we present results of numerical integration of the system of equations (\ref{spinevoleq})-(\ref{modeevoleq}) and compare with the approximate analytic solution derived in section \ref{approxanalytic} above. For realistic accretion rates, numerical integration of (\ref{spinevoleq})-(\ref{modeevoleq}) is challenging because of a drastic separation of timescales. The timescale of accretion-induced spin-up is of order $\sim 10^{-3} \,{\rm yr}^{-1}$, while the characteristic timescale of the mode amplitude equation away from resonance is $\sim \Omega^{-1} \sim 10^{-2} \, {\rm s}^{-1}$. Stable numerical integration of system (\ref{spinevoleq})-(\ref{modeevoleq}) using a Runge-Kutta scheme from $\Omega = 0$ to $\Omega = \Omega_{\rm res}$ takes a very long time and is not a desirable approach. What we do instead is start numerical integration close to resonance and use initial conditions for mode amplitude derived from the approximate solution (\ref{cAapproxI}) to equation of motion (\ref{modeampEOMII}). In figure \ref{fig:omega} we show the time evolution of the spin frequency of the accretor obtained from integrating (\ref{spinevoleq})-(\ref{modeevoleq}) using such approximate initial conditions. The resonant mode used here is the $l=5,m=3, w=-0.5266$ generalized {\it r}-mode, and we used the same values for tidal ratio and mass ratio as reported in table \ref{table:numericalvalues}. We normalize units of time $\Omega_{\rm res} = 1$. In these units we used a value of $\dot{\Omega}_{\rm acc} = 10^{-9}$, its value in Hz/yr depending on the value of the orbital frequency. For an orbital period of $\sim 5$ minutes, the accretion spin-up rate is $\sim 3 \times 10^{-5} \, {\rm Hz/yr}$. For these parameters bound (\ref{barnubound}) is
\be\label{53bound}
\frac{\bar{\nu}_{53}}{\Omega^{\rm res}_{53}} \gg 2.2 \times 10^{-2},
\ee
which is satisfied as one can see from table \ref{table:numericalvalues}. We thus expect the mode to stabilize the spin for this particular choice of parameters.

\begin{figure}
\hspace*{\fill}
\includegraphics[width=5.7cm,angle=270]{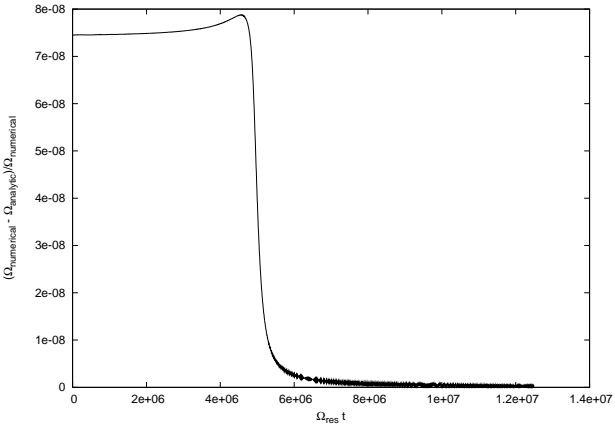}
\hspace*{\fill}
\caption{This figure shows the relative error between the spin frequency obtained numerically and the approximate analytic solution derived in subsection \ref{approxanalytic}.}\label{fig:error}
\end{figure}

In figure \ref{fig:error} we show the relative error between the spin frequency obtained from numerical integration of (\ref{spinevoleq})-(\ref{modeevoleq}) and the full analytic approximation (\ref{fullsol}). The relative error is less than $10^{-7}$, which gives additional strong justification to the various approximations made in deriving this analytic solution. We attribute the initial discrepancy between the numerical solution and the analytic solution to our practical choice of initial conditions for numerical integration. 

\subsection{Lifetime of synchronizer}\label{sec:lifetime}

The synchronizing mechanism we propose is based on exciting a mode close to its resonance frequency over a long period of time. This process obviously cannot last forever since the mode amplitude grows as $\sqrt{t - t_0}$ in the regime where the spin is stabilized, as can be seen by combining (\ref{Omegalarget}) and (\ref{spinevoleqII})
\bea
|\gamma_\alpha|^2 &=& |\gamma_\alpha(0)|^2 + \frac{\dot{\Omega}_{\rm acc} t  - \Omega_\alpha^{\rm res}}{\nu_\alpha} + \frac{1}{\sqrt{\lambda_\alpha^2\nu_\alpha\dot{\Omega}_{\rm acc}(t-t_0)}}  \nn\\
&\simeq& \frac{\dot{\Omega}_{\rm acc} t  - \Omega_\alpha^{\rm res}}{\nu_\alpha} \simeq\frac{\dot{\Omega}_{\rm acc}}{\nu_\alpha}(t - t_0).
\eea
After a certain amount of time the mode amplitude will reach a value where the resonant mode couples significantly to other modes of the star through non-linear hydrodynamical effects (see \cite{Schenk,jb1,jb2}), which saturates the growth of the mode amplitude. As shown by \cite{jb1,jb2}, a precise understanding of the evolution of a perturbed star in the regime where a mode saturates requires a simulation involving thousands of modes. 

Denoting the saturation amplitude $c_{\rm sat}$, we obtain the following estimate for the duration $T_{\rm lif}$ of the stabilization mechanism presented
\bea
T_{\rm lif} &\sim&  |c_{\rm sat}|^2 \frac{\nu_\alpha}{|\mathcal{F}_\alpha|^2\dot{\Omega}_{\rm acc}} \sim |c_{\rm sat}|^2\frac{\bar{\nu}_\alpha}{\dot{\Omega}_{\rm acc}}\nn \\
&\sim& 1.5 \times 10^{10} \frac{|c_{\rm sat}|^2}{\Omega_{\rm res}},  \label{lifetimeestimate}
\eea
for the $l=5,m=3$ {\it r}-mode and the same model parameters as used to generate figure \ref{fig:omega}. The lifetime of the linear synchronizer\footnote{By "linear synchronizer" we here mean the regime where the mode amplitude is small enough so that it can be evolved using the linear, leading-order mode amplitude evolution equation, neglecting non-linear mode-mode couplings. However the coupling between the tides and the uniform (rigid) rotation mode is still non-linear (quadratic in the mode amplitude). } thus depends crucially on the mode saturation amplitude and could be as short as a few orbits but also as long as millions of years. For example an optimistic saturation amplitude of $\sim 10^{-2}$ and a slow accretion rate, say $10^{-7}M_\odot / {\rm yr}$ yields a lifetime of order $10^6$ years. For sake of comparison, the radiation reaction timescale of a binary with components of $0.6\, M_\odot$ and $0.2\, M_\odot$ and orbital period of $10$ minutes is $t_{\rm rr} \sim 8 \times 10^7$ years. After the resonant mode saturates, the fate of the binary remains an open problem.

\subsection{Polytropic stellar model}

Clearly the incompressible model we used above is not very realistic. It is therefore worthwhile checking if changes in mode eigenfunctions due to using a more realistic stellar model affect significantly the strength of tidal coupling and spin back-reaction parameters of our candidate synchronizers. Here we report values for mode frequencies, tidal coupling and back-reaction parameters computed from a polytropic star of index $n=3/2$ (or equivalently $\Gamma_1 = 5/3$) and spinning at frequency $\Omega^2 = 0.01 M/R^3$. There is no buoyancy in the star, so that {\it g}-modes are degenerate, zero-frequency modes. For this stellar model, the mode frequencies, tidal coupling and back-reaction parameters are, for the two main candidate modes discussed previously, 
\be
\frac{(w_{l=4,m=2})_{n=3/2}}{(w_{l=4,m=2})_{\rm incomp}} = 0.861,
\ee
\be
\frac{(w_{l=5,m=3})_{n=3/2}}{(w_{l=5,m=3})_{\rm incomp}} = 0.823,
\ee
\be
\frac{|\mathcal{F}_{l=4,m=2}|_{n=3/2}}{|\mathcal{F}_{l=4,m=2}|_{\rm incomp}} = 0.0897,
\ee
\be
\frac{|\mathcal{F}_{l=5,m=3}|_{n=3/2}}{|\mathcal{F}_{l=5,m=3}|_{\rm incomp}} = 0.0700 \label{f53n32} 
\ee
and
\be
\frac{(\bar{\nu}_{l=4,m=2})_{n=3/2}}{(\bar{\nu}_{l=4,m=2})_{\rm incomp}} = 1.94,
\ee
\be
\frac{(\bar{\nu}_{l=5,m=3})_{n=3/2}}{(\bar{\nu}_{l=5,m=3})_{\rm incomp}} =  1.65. \label{newbarnu}
\ee
For the $l=5,m=3$ mode, the suppression factor in tidal coupling strength [cf. Eq.(\ref{f53n32})] is such that criterion (\ref{barnubound}) is not cleanly satisfied anymore. For an accretion rate $\dot{\Omega}_{\rm acc} = 10^{-9}$ (in units of time where $\Omega_{53}^{\rm res} = 1$), which still corresponds to a mass transfer rate of order $10^{-5} M_\odot \,\, {\rm yr}^{-1}$, bound (\ref{barnubound}) becomes
\be\label{newbarnubound}
\frac{\bar{\nu}_{53}}{\Omega_{53}^{\rm res}} \gg 4.7,
\ee 
From (\ref{newbarnu}) and table \ref{table:numericalvalues} we have
\be
\frac{\bar{\nu}_{53}}{\Omega_{53}^{\rm res}} = 24.02,
\ee
which is only a factor of 5 larger than the critical value given in (\ref{newbarnubound}). Therefore one cannot say whether or not this mode will act as a synchronizer generically, since slight changes in accretion rate or tidal ratio may push the mode on one side or another of the bound. For example tidal coupling strength squared for an $l=5$ mode scales with tidal ratio as $(R_1/d)^{12}$. Thus a slight change in tidal ratio may easily increase or decrease the critical value in bound (\ref{newbarnubound}) by an order of magnitude. 

Lastly the lifetime of the synchronizer is still given by the following expression
\be
T_{\rm lif} \sim |c_{\rm sat}|^2\frac{\bar{\nu}_\alpha}{\dot{\Omega}_{\rm acc}},
\ee
so for the same accretion rate, the lifetime increases only by a factor of 1.65 [cf. Eq.(\ref{newbarnu})] compared to the incompressible case. 

\section{Conclusion and future work}

In this paper we suggest a non-dissipative mechanism that can stabilize the spin of the accretor in an ultra-compact binary white dwarf system. The effect that stabilizes the spin of the accreting white dwarf is the back-reaction of a resonantly driven generalized {\it r}-mode on the uniform rotation mode. For the model we analyze in detail we assume that only a single mode is dynamically relevant (see section \ref{models} for more details). We also assume slow rotation of the accretor, constant accretion torque, constant orbital parameters and a perfect incompressible fluid equation of state. By integrating the equations of motion of our model both numerically and analytically we show that pure hydrodynamics may stabilize the spin of the accretor. No dissipative effects are required for synchronization. However the lifetime of the synchronizing mechanism is limited since it requires driving the mode continuously very close to resonance. Eventually the mode enters a non-linear regime where it couples to other modes, which saturates the growth of the resonant mode's amplitude. At that point the synchronizing mechanism most likely shuts off. A correct computation of the saturation amplitude involves simulating a network of coupled modes following the footsteps of \cite{jb1,jb2}, a challenging task that we leave here as an open problem. More future work includes solving the full equations of motion, where the accretion rate and orbital parameters are computed and evolved rather than assumed to be constant external input quantities. In addition if mode damping timescales turn out to be comparable to the lifetime of the synchronizer, then one should include damping when solving the mode amplitude evolution equation. It would also be interesting to investigate the properties of the stabilization mechanism if the accretor experiences core crystallization, as the properties of the {\it r}-modes (i.e. frequency spectrum, strength of tidal coupling, etc...) existing in the fluid shell surrounding the core differ from the case of a fluid spheroid.

\section*{acknowledgements}
\'{E}.R. wishes to thank Jeandrew Brink for many useful discussions on {\it r}-modes. \'{E}.R. and E.S.P. acknowledge support from NASA ATP grant NNG04GK98G awarded to E.S.P.

\appendix

\section{Perturbations of uniformly rotating stars}\label{app:perttheory}

In this appendix we review basic material from perturbation theory of uniformly rotating stars developed in \cite{Schenk} that is required for the analysis of this paper. This appendix is rather technical but the formalism reviewed here is essential in correctly analyzing the tidal response of a rotating star. We hope to provide the interested reader an accessible introduction to the beautiful, albeit mathematically heavy, work of \cite{Schenk}. We refer the reader to that paper for complete details. We employ units where $G=c=1$ throughout. 

Begin by considering an unperturbed star of mass $M$ and radius $R$ uniformly rotating at constant spin frequency $\Omega$. The radius of the spinning star is defined as the radius of the sphere enclosing the same total volume as the star. The hydrodynamical equations of motion in the frame corotating with the star are
\be
\frac{\partial \rho}{\partial t} + \bmath{\nabla}\cdot(\rho \bmath{u}) = 0, 
\ee
\be
\frac{D \bmath{u}}{D t}  + \bmath{\Omega}\times(2\bmath{u} + \bmath{\Omega} \times \bmath{x}) = -\frac{1}{\rho} \bmath{\nabla} p - \bmath{\nabla}\Phi + \bmath{a}_{\rm ext},
\ee
where $\rho$ is mass density, $p$ is pressure, $\bmath{u}$ is the fluid velocity, $\Phi$ the self-gravitational potential and $\bmath{a}_{\rm ext}$ is an acceleration field produced by some external perturbation. The operator $D/Dt$ is the usual convective time derivative. For the unperturbed star, $\rho$ does not depend on time and the velocity $\bmath{u}$ vanishes. The corotating frame is related to the inertial frame by the simple coordinate transformation $\phi = \varphi - \int \Omega dt $, where $\phi$ is the corotating azimuthal coordinate and $\varphi$ is the inertial azimuthal coordinate. If a fluid element of the unperturbed star located at $\bmath{x}$ is displaced by the vector $\bmath{\xi}(\bmath{x},t)$, the linearized hydrodynamical equations for the Lagrangian displacement vector $\bmath{\xi}$ take the form
\be\label{pertevol}
\ddot{\bmath{\xi}} + \bmath{B}\cdot\dot{\bmath{\xi}} + \bmath{C}\cdot\bmath{\xi} = \bmath{a}_{\rm ext},
\ee
where the action of the operator $\bmath{B}$ on a vector is defined as
\be
\bmath{B}\cdot\bmath{\xi} \equiv 2\bmath{\Omega} \times \bmath{\xi}.
\ee
For stars without buoyancy, the action of the operator $\bmath{C}$ on a vector is given by
\be
\bmath{C}\cdot\bmath{\xi} = \bmath{\nabla}\left[\frac{\Gamma p}{\rho^2}\delta \rho + \delta \Phi \right] = \bmath{\nabla}\left[\frac{1}{\rho}\delta p + \delta \Phi \right],
\ee
where $\Gamma$ is the adiabatic index of the perturbation, which we here take to be the same as the background star. The quantities $\delta p$ and $\delta \rho$ are the Eulerian perturbations in the pressure and density respectively. In terms of the displacement vector the density perturbation is
\be
\delta \rho = - \bmath{\nabla} \cdot(\rho \bmath{\xi}).
\ee
The gravitational potential perturbation $\delta \Phi$ obeys
\be
\nabla^2 \delta \Phi = 4\pi \delta \rho.
\ee
The fluid displacement equation of motion (\ref{pertevol}) can also be obtained from the following Lagrangian density

\be\label{Lagrangian}
\mathcal{L} = \frac{1}{2}\dot{\bmath{\xi}}\cdot\dot{\bmath{\xi}} + \frac{1}{2}\dot{\bmath{\xi}}\cdot\bmath{B} \cdot \bmath{\xi} - \frac{1}{2}\bmath{\xi}\cdot\bmath{C}\cdot\bmath{\xi} + \bmath{a}_{\rm ext} \cdot \bmath{\xi}.
\ee
The normal modes of the star are obtained by assuming displacements of the form $\bmath{\xi}(\bmath{x},t) = e^{-i\omega t} \bmath{\xi}(\bmath{x})$ and no external forces. One must then solve the following eigenvalue equation for the normal modes
\be\label{eigenvalue}
\Big[ -\omega^2 - i\omega\bmath{B} + \bmath{C} \Big]\cdot\bmath{\xi}(\bmath{x}) = 0.
\ee
A mode $(\bmath{\xi}_A,\omega_A)$ is solution to Eq.(\ref{eigenvalue}). Later in the paper we will use the notation $w_A \equiv \omega_A / 2\Omega$ for dimensionless mode frequencies. Note that given a mode $(\bmath{\xi}_A,\omega_A)$ with $\omega_A \neq 0$, there exists another distinct mode $(\bmath{\xi}_B, \omega_B)$ solution to (\ref{eigenvalue}) with $\bmath{\xi}_B = \bmath{\xi}_A^\ast$ and $\omega_B = -\omega_A$. Thus modes with non-zero frequency always occur in pairs. 

\subsection{Phase space expansion formalism}

The eigenvalue equation (\ref{eigenvalue}) satisfied by a given mode is peculiar in the sense that the operator acting on the perturbation depends on the eigenvalue $\omega$. This property leads to the fact that although it is possible to find a complete set of solutions to (\ref{eigenvalue}) so that a generic fluid displacement may be expanded as
\be\label{displacementcoordexp}
\bmath{\xi}(\bmath{x},t) = \sum_\alpha q_\alpha(t) \bmath{\xi}_\alpha(\bmath{x}),
\ee
the evolution equations for the mode amplitude coefficients $q_\alpha(t)$ are not in general uncoupled from one another. To get around this problem one needs to look at the fluid perturbation in phase space (\cite{Schenk,dysonschutz}). A general fluid displacement vector is expanded in phase space as follows 
\be\label{phasespaceexpansion}
\bmath{\zeta}(\bmath{x},t) \equiv \left[\begin{array}{c} \bmath{\xi}(\bmath{x},t) \\ \bmath{\pi}(\bmath{x},t) \end{array} \right] = \sum_A c_A(t) \left[\begin{array}{c} \bmath{\xi}_A(\bmath{x}) \\ \bmath{\pi}_A(\bmath{x}) \end{array} \right],
\ee
where 
\be
\bmath{\pi} = \dot{\bmath{\xi}} + \frac{1}{2}\bmath{B}\cdot\bmath{\xi}
\ee
is the momentum conjugate to $\bmath{\xi}$ obtained from the Lagrangian density (\ref{Lagrangian}). Using (\ref{pertevol}) the evolution equation for the phase space vector $\bmath{\zeta}$ is easily seen to be
\be\label{phasespacepertevol}
\dot{\bmath{\zeta}} = \left(\begin{array}{cc} -\frac{1}{2}\bmath{B} & \bmath{1} \\ - \bmath{C} + \frac{1}{4}\bmath{B}^2 & -\frac{1}{2}\bmath{B} \end{array}\right) \cdot\bmath{\zeta} + \left[\begin{array}{c} 0 \\ \bmath{a}_{\rm ext} \end{array} \right] \equiv \bmath{T}\cdot \bmath{\zeta} + \bmath{F}.
\ee
Thus the eigenvalue equation equivalent to (\ref{eigenvalue}) in phase space is
\be\label{zetaA}
\left[\bmath{T} + i\omega_A \right] \bmath{\zeta}_A = 0.
\ee
An important point to note here is that operator $\bmath{T}$ is not Hermitian, which leads to the existence of left eigenvectors $\bmath{\chi}_A$ distinct from the right eigenvectors $\bmath{\zeta}_A$ (which incidentally do {\it not} form a complete basis; to form a complete basis, one needs to include all Jordan chain modes [see Appendix A of \cite{Schenk} for a complete discussion]) that satisfy
\be\label{chiA}
\left[\bmath{T}^\dagger - i\omega_A^\ast\right] \bmath{\chi}_A = 0.
\ee
We will write the right and left eigenmodes as follows
\bea
\bmath{\zeta}_A = \left[\begin{array}{c} \bmath{\xi}_A \\ \bmath{\pi}_A \end{array} \right], \label{eigenmodesdefA}\\
\bmath{\chi}_A = \left[\begin{array}{c} \bmath{\sigma}_A \\ \bmath{\tau}_A \end{array}\right].\label{eigenmodesdefB}
\eea
Now it is always possible to choose $\bmath{\chi}_A$ to be dual to $\bmath{\zeta}_A$ in the sense that
\be\label{phasespaceorthon}
\langle \bmath{\chi}_A, \bmath{\zeta}_B \rangle \equiv \langle \bmath{\sigma}_A, \bmath{\xi}_B \rangle + \langle \bmath{\tau}_A, \bmath{\pi}_B\rangle = \delta_{AB},
\ee
where the inner product of two vector fields is defined as
\be
\langle \bmath{\xi}, \bmath{\xi}^\prime \rangle = \int d^3x \, \rho(\bmath{x})\, \bmath{\xi}^\ast(\bmath{x}) \cdot \bmath{\xi}^\prime(\bmath{x}).
\ee
Substituting (\ref{eigenmodesdefA})-(\ref{eigenmodesdefB}) into (\ref{zetaA}) and (\ref{chiA}) yields
\bea
\bmath{\pi}_A &=& -i\omega_A\bmath{\xi}_A + \frac{1}{2}\bmath{B}\cdot\bmath{\xi}_A, \label{kinematicidsA}\\
\bmath{\sigma}_A &=& i\omega_A^\ast \bmath{\tau}_A - \frac{1}{2}\bmath{B}\cdot\bmath{\tau}_A. \label{kinematicidsB}
\eea

If we now assume that all mode frequencies are real, it is then possible to show the following (this proof is a little involved; we refer the reader to Appendix A of \cite{Schenk} for details) 
\be\label{tauA}
\bmath{\tau}_A = -\frac{i}{b_A} \bmath{\xi}_A,
\ee
where $b_A$ is a (real) constant. This constant is determined from (\ref{phasespaceorthon}), which may be rewritten as 
\be\label{bA}
b_A \delta_{AB} = \langle \bmath{\xi}_A, i\bmath{B}\cdot\bmath{\xi}_B \rangle + (\omega_A + \omega_B)\langle \bmath{\xi}_A,\bmath{\xi}_B \rangle.
\ee
The orthonormality relation (\ref{phasespaceorthon}) allows us to invert expansion (\ref{phasespaceexpansion}) to get
\be\label{cAdef}
c_A = \frac{i}{b_A} \langle \bmath{\xi}_A, -i\omega_A\bmath{\xi} + \frac{1}{2}\bmath{B}\cdot\bmath{\xi} + \bmath{\pi} \rangle.
\ee
Finally, (\ref{phasespaceorthon}) may also be used to extract from (\ref{phasespacepertevol}) the following evolution equation for mode amplitude
\be\label{modeevolution}
\dot{c}_A + i\omega_A c_A = \frac{i}{b_A}\langle \bmath{\xi}_A, \bmath{a}_{\rm ext} \rangle.
\ee
To summarize, the tidal response of a rotating star is described by a Lagrangian fluid displacement vector field $\bmath{\xi}(\bmath{x},t)$ and its canonically conjugate momentum $\bmath{\pi}(\bmath{x},t)$. These two vector fields may be decomposed in phase space into normal modes according to (\ref{phasespaceexpansion}), a given normal mode $(\bmath{\xi}_A,\omega_A)$ satisfying the eigenvalue equation (\ref{eigenvalue}). The time evolution of a given mode amplitude is governed by Eq.(\ref{modeevolution}).

\subsection{An important set of zero-frequency Jordan-chain modes}\label{sec:Jordanchains}

As mentioned in the previous section the modes $\bmath{\zeta}_A$ analyzed above do not form a complete basis on which one may expand an arbitrary fluid perturbation. To form a complete basis one needs to add all Jordan-chain modes. A Jordan chain of length $p_A$ associated with a given eigenfrequency $\omega_A$ is a set of vectors $\bmath{\zeta}_{A,\sigma}$, $0 \leq \sigma \leq p_A$ satisfying
\be\label{Jordan1}
\left[\bmath{T} + i\omega_A\right]\bmath{\zeta}_{A,\sigma} = \bmath{\zeta}_{A,\sigma-1},
\ee
for $1 \leq \sigma \leq p_A$ and
\be\label{Jordan2}
\left[\bmath{T} + i\omega_A\right]\bmath{\zeta}_{A,0} = 0.
\ee
The associated left eigenvectors satisfy
\be\label{Jordan3}
\left[\bmath{T}^\dagger - i\omega_A^\ast\right]\bmath{\chi}_{A,\sigma} = \bmath{\chi}_{A,\sigma-1},
\ee
for $1 \leq \sigma \leq p_A$ and
\be\label{Jordan4}
\left[\bmath{T}^\dagger - i\omega_A^\ast\right]\bmath{\chi}_{A,0} = 0.
\ee
The generalized orthonormality condition for Jordan chains adopted in \cite{Schenk} is
\be\label{Jordannorm}
\langle \bmath{\chi}_{A,\sigma},\bmath{\zeta}_{B,\tau} \rangle = \delta_{AB}\delta_{p_A,\sigma + \tau}
\ee
The equations of motion for the associated mode amplitudes in the absence of external forcing can be shown to be (see Appendix A of \cite{Schenk})
\be
\dot{c}_{A,\sigma} + i\omega_Ac_{A,\sigma} = c_{A,\sigma+1}, \,\, {\rm for}\,\, \sigma < p_A \label{JordanchainEOMa} 
\ee
\be
\dot{c}_{A,p_A} + i\omega_Ac_{A,p_A} = 0. \label{JordanchainEOMb}
\ee

The main reason motivating us to discuss Jordan chains in this paper is that all differential rotation modes must be Jordan chains of length one. One proves this as follows. The fluid velocity perturbation corresponding to a differential rotation mode is of the form
\be
\frac{\partial \bmath{\xi}(\bmath{x},t)}{\partial t} \equiv  \delta \bmath{u}(\bmath{x}) = r_\perp \delta \Omega(r_\perp) \hat{\bmath{e}}_\phi,
\ee
where $r_\perp = r\sin\theta$. The velocity perturbation being time-independent first proves that the mode is a zero-frequency perturbation. Now any normal mode is a Jordan chain of length $p \geq 0$, so we have (dropping $A$ subscripts)
\be\label{uofxi}
\delta \bmath{u}(\bmath{x}) = \sum_{\sigma = 0}^p \dot{c}_\sigma(t) \bmath{\xi}_\sigma(\bmath{x}).
\ee
The mode amplitude equations of motion (\ref{JordanchainEOMa})-(\ref{JordanchainEOMb}) show that $c_\sigma(t)$, in the case of a zero-frequency chain, is a polynomial in $t$ of order $p-\sigma$. Since $\delta \bmath{u}$ is independent of time but non-vanishing, then the chain must be of length one. In particular, the uniform rotation mode ($\delta\Omega(r_\perp) = \delta\Omega = {\rm constant}$) which moves the star from a given background spin frequency $\Omega$ to a new background spin frequency $\Omega + \delta \Omega$ falls into this category. From Eq.(\ref{uofxi}) one immediately finds that for a generic differential rotation mode, the mode function $\bmath{\xi}_0$ is 
\be
\bmath{\xi}_0 = \gamma\,\delta \bmath{u} ,
\ee
where $\gamma$ is a constant fixed by an overall mode normalization condition. The defining equations of Jordan chains (\ref{Jordan1}) and (\ref{Jordan2}) are then used to show that $\bmath{\xi}_1$ is solution to
\be\label{urxi1}
\bmath{C}\cdot \bmath{\xi}_1 = - \bmath{B}\cdot\bmath{\xi}_0.
\ee
In the case of uniform rotation, we use the convention
\be\label{urxi0}
\bmath{\xi}_0 = \hat{\bmath{\Omega}}\times \bmath{x}.
\ee
The complete expression for the uniform rotation Jordan chain is therefore
\be
\bmath{\zeta}_0 = \left[\begin{array}{c} \bmath{\xi}_0 \\ \frac{1}{2}\bmath{B}\cdot\bmath{\xi}_0 \end{array}\right], \label{zeta0ur} 
\ee
\be
\bmath{\zeta}_1 = \left[\begin{array}{c} \bmath{\xi}_1 \\ \frac{1}{2}\bmath{B}\cdot\bmath{\xi}_1 + \bmath{\xi}_0 \end{array}\right], \label{zeta1ur}
\ee
with $\bmath{\xi}_0$ and $\bmath{\xi}_1$ given by (\ref{urxi0}) and (\ref{urxi1}) respectively. Throughout this paper, when we say the spin of the star changes, we mean that the amplitude of the above uniform rotation Jordan chain mode changes. The amplitude of the uniform rotation mode defines the natural coordinate system corotating with the star in which stellar perturbation theory is applied. In the specific application we consider in this paper, it is the time dependence of the coordinate system corotating with the star due to the evolution of the uniform rotation mode that is responsible for the resonant excitation of normal modes in an accreting white dwarf part of a binary system.

Lastly, it will be useful later on to know the left Jordan chain associated with the uniform rotation mode. The defining equations (\ref{Jordan3}) and (\ref{Jordan4}), along with (\ref{Jordannorm}) and (\ref{zeta0ur})-(\ref{zeta1ur}) give 
\be\bmath{\chi}_0 = \frac{1}{\beta}\left[\begin{array}{c} -\frac{1}{2}\bmath{B}\cdot\bmath{\xi}_0 \\ \bmath{\xi}_0 \end{array}\right], \label{chi0ur} 
\ee
\be
\bmath{\chi}_1 = \frac{1}{\beta}\left[\begin{array}{c} \frac{1}{2}\bmath{B}\cdot\bmath{\xi}_1 + \bmath{\xi}_0  \\ -\bmath{\xi}_1 \end{array}\right], \label{chi1ur}
\ee
where
\be\label{betadef}
\beta = \langle \bmath{\xi}_0, \bmath{\xi}_0 \rangle  + \langle \bmath{\xi}_1 , \bmath{C} \cdot \bmath{\xi}_1 \rangle
\ee
The constant $\beta$ ensures that the left and right eigenvectors given in (\ref{chi0ur})-(\ref{chi1ur}) and (\ref{zeta0ur})-(\ref{zeta1ur}) satisfy the orthonormality condition (\ref{Jordannorm}).
 
\subsection{Mode pair index notation}

As mentioned previously, modes with non-zero frequencies always occur in pairs. We may then rewrite the phase space expansion index $A$ as the pair of indices $(\alpha, \epsilon)$, $\epsilon = \pm$, with the following identifications
\be
\omega_{\alpha,+} = \omega_\alpha  \label{modepairnotationA}
\ee
\be
\omega_{\alpha,-} = - \omega_\alpha \label{modepairnotationB}
\ee
\be
\bmath{\xi}_{\alpha,+} = \bmath{\xi}_\alpha \label{modepairnotationC}
\ee
\be
\bmath{\xi}_{\alpha,-} = \bmath{\xi}_\alpha^\ast \label{modepairnotationD}
\ee
Equations (\ref{modepairnotationA}) to (\ref{modepairnotationD}) and (\ref{bA}) then imply 
\be
b_{\alpha,+} =  b_{\alpha} \label{balphaA}
\ee
\be
b_{\alpha,-} = -b_{\alpha}. \label{balphaB}
\ee
Since the fluid displacement and its conjugate momentum are real vector fields, Eqs.(\ref{modepairnotationA}) to (\ref{modepairnotationD}), (\ref{cAdef}) and (\ref{balphaA})-(\ref{balphaB}) yield
\be
c_{\alpha,+} = c_{\alpha} \label{calphaA}
\ee
\be
c_{\alpha,-} = c_{\alpha}^\ast \label{calphaB}
\ee
If the perturbation is dominated by modes with non-zero frequency we may rewrite expansion (\ref{phasespaceexpansion}) as follows
\bea
 \left[\begin{array}{c} \bmath{\xi}(\bmath{x},t) \\ \bmath{\pi}(\bmath{x},t) \end{array} \right] &=& \sum_A c_A(t) \left[\begin{array}{c} \bmath{\xi}_A(\bmath{x}) \\ \bmath{\pi}_A(\bmath{x}) \end{array} \right] \nn \\
&=& \sum_{\alpha} c_\alpha(t)\left[\begin{array}{c} \bmath{\xi}_\alpha(\bmath{x}) \\ \bmath{\pi}_\alpha(\bmath{x}) \end{array} \right] + \,\, {\rm c.c.},
\eea
where ${\rm c.c.}$ stands for complex conjugate.

\section{Detailed derivation of spin frequency evolution equation}\label{app:spinevol}

The differential equation governing the time evolution of the spin frequency $\bmath{\Omega}$ of the star can be obtained using results of \cite{Schenk}, who rigorously derive the evolution equation of the uniform rotation mode amplitude to second order in stellar perturbation theory. Here we shall instead derive the evolution equation for the spin frequency following the procedure sketched in the introduction. Our evolution equation of course matches the result of \cite{Schenk}. However we believe our derivation is slightly simpler to follow so it is worth including it here. We start from the expression for the component of the angular momentum of the perturbed star along its unperturbed spin axis, and expand it to second order in the perturbation. A direct computation yields  
\bea
J_{\rm star} &=& \int d^3x \, \rho(\bmath{x})\,\hat{\bmath{\Omega}} \cdot\Big\{ (\bmath{x} + \bmath{\xi}) \times \Big[\bmath{\Omega}_0 \times (\bmath{x} + \bmath{\xi}) + \dot{\bmath{\xi}}\Big]\Big\} \nn \\
&=& \int d^3x \, \rho(\bmath{x})\,\hat{\bmath{\Omega}} \cdot\Big\{ (\bmath{x} + \bmath{\xi}) \times \Big[\bmath{\Omega}_0 \times \bmath{x} + \bmath{\pi} \Big]\Big\} \nn \\
&=& \int d^3x \, \rho \, \Omega r_\perp^2  + \int d^3x \, \rho \, \hat{\bmath{\Omega}}\cdot(\bmath{\xi} \times \bmath{\pi}) \nn \\
&& +  \int d^3x \, \rho \Big[ (\bmath{\Omega}_0 \times \bmath{\xi})\cdot(\hat{\bmath{\Omega}} \times \bmath{x}) + \hat{\bmath{\Omega}}\cdot(\bmath{x}\times\bmath{\pi})\Big] \nn \\ 
&\equiv& \Omega_0 I_0 + J_2[\bmath{\zeta},\bmath{\zeta}] + J_1[\bmath{\zeta}], \label{Jstarexpanded}
\eea
where $\bmath{\Omega}_0 = \Omega_0 \hat{\bmath{\Omega}}$ is the unperturbed spin frequency vector, which was denoted by $\bmath{\Omega}$ in Appendix \ref{app:perttheory}. The full spin frequency of the star is now written as 

\be
\bmath{\Omega} = \bmath{\Omega}_0 + \delta \Omega\, \hat{\bmath{\Omega}}.
\ee
Let us now focus on the term $J_1[\bmath{\zeta}]$. Using the defining equations (\ref{urxi0}) and (\ref{chi0ur}) for the uniform rotation Jordan chain, we may rewrite $J_1[\bmath{\zeta}]$ as follows 
\bea
J_1[\bmath{\zeta}] &=& \int d^3x \, \rho \Big[ (\bmath{\Omega}_0 \times \bmath{\xi})\cdot(\hat{\bmath{\Omega}} \times \bmath{x}) + \hat{\bmath{\Omega}}\cdot(\bmath{x}\times\bmath{\pi})\Big] \nn \\
&=& \int d^3x \, \rho \Big[-\bmath{\xi}\cdot(\bmath{\Omega}_0 \times \bmath{\xi}_0) + \bmath{\pi}\cdot\bmath{\xi}_0\Big] \nn \\
&=& \langle -\frac{1}{2}\bmath{B}\cdot\bmath{\xi}_0, \bmath{\xi} \rangle + \langle \bmath{\xi}_0, \bmath{\pi} \rangle \nn \\
&=& \beta \langle \bmath{\chi}_0 , \bmath{\zeta} \rangle \nn \\
&=& \beta \, c_1(t),
\eea
where the last line has been obtained using Jordan-chain orthonormality condition (\ref{Jordannorm}). Thus the piece of the angular momentum linear in the perturbation depends only on the amplitude $c_1$ of the uniform rotation mode. It is independent of all other modes of the star from orthonormality condition (\ref{Jordannorm}). It remains to relate the coefficient $c_1$ to the spin frequency perturbation and to compute the quantity $\beta$ defined in (\ref{betadef}) . Using (\ref{zeta0ur}), the velocity perturbation is 
\be
\delta \bmath{u} = r_\perp \delta \Omega \,\hat{\bmath{e}}_\phi =  \delta \Omega \,(\hat{\bmath{\Omega}}\times\bmath{x}) = \delta \Omega\, \bmath{\xi}_0,
\ee
which gives simply [cf. Eq.(\ref{uofxi})]
\be
\dot{c}_0 = c_1 =  \delta \Omega. 
\ee
We now turn to the computation of $\beta$ [cf. Eq.(\ref{betadef})]. The first term of the right-hand side of (\ref{betadef}) is easily shown to be
\be
\langle \bmath{\xi}_0 , \bmath{\xi}_0 \rangle = I_0,
\ee
the unperturbed moment of inertia of the star about its spin axis. This result leads us to expect that the second term of (\ref{betadef}) is related to the perturbation of the star's moment of inertia under uniform spin-up. The moment of inertia of the perturbed star about its spin axis is given by
\be
I = \int d^3x \, \rho \, \left\{ (\bmath{x} + \bmath{\xi})^2 - [(\bmath{x} + \bmath{\xi})\cdot\hat{\bmath{\Omega}}]^2\right\}.
\ee
Expanding this expression to leading order in $\bmath{\xi}$ yields after some algebra and an integration by parts
\be
I = I_0 - \int d^3x \,  \bmath{\nabla}\cdot(\rho \bmath{\xi}) [\bmath{x}^2 - (\bmath{x}\cdot\hat{\bmath{\Omega}})^2]
\ee
If the perturbation consists of uniform spin-up, we then have, since $\rho \bmath{\xi}_0$ is divergenceless, 
\be\label{deltaIdef}
I = I_0 - c_1\int d^3x \,  \bmath{\nabla}\cdot(\rho \bmath{\xi}_1) [\bmath{x}^2 - (\bmath{x}\cdot\hat{\bmath{\Omega}})^2] \equiv I_0 + \delta I.
\ee
On the other hand we have, from (\ref{urxi1})
\bea
\langle \bmath{\xi}_1, \bmath{C}\cdot\bmath{\xi}_1 \rangle &=& -\langle \bmath{\xi}_1, \bmath{B}\cdot\bmath{\xi}_0 \rangle \nn \\
&=&  - \int d^3x \, \rho \bmath{\xi}_1 \cdot \big[2\bmath{\Omega}_0 \times (\hat{\bmath{\Omega}} \times\bmath{x}) \big] \nn \\
&=& \Omega_0 \int d^3x \, \rho \bmath{\xi}_1 \cdot \bmath{\nabla}\big[\bmath{x}^2 - (\hat{\bmath{\Omega}}\cdot\bmath{x})^2 \big] \nn \\
&=& -\Omega_0 \int d^3x \, \bmath{\nabla}\cdot(\rho \bmath{\xi}_1) \big[\bmath{x}^2 - (\hat{\bmath{\Omega}}\cdot\bmath{x})^2 \big] \nn \\
&=& \frac{\Omega_0}{c_1}\delta I,
\eea
where the last line has been obtained using (\ref{deltaIdef}). We finally have
\bea
J_1[\bmath{\zeta}] &=& \left[I_0 + \frac{\Omega_0}{c_1} \delta I\right]c_1 \nn \\
&=& I_0 \delta \Omega + \Omega_0 \delta I \nn \\
&=& \left[1 + \frac{d\ln I_0}{d\ln \Omega_0}\right]I_0 \delta \Omega.
\eea
In this paper we will assume that the star spins slowly enough so that the change in moment of inertia due to change in spin may be neglected. Taking a time derivative of (\ref{Jstarexpanded}) then yields, in the context of the adiabatic approximation detailed in section \ref{sec:adiab},
\be\label{dotOmegafinal}
\frac{d\Omega}{dt} = \frac{1}{I}\left[\mathcal{T}_{\rm ext} - \frac{d}{dt}\int d^3x \, \rho\, \hat{\bmath{\Omega}}\cdot(\bmath{\xi} \times \bmath{\pi}) \right],
\ee
where $\mathcal{T}_{\rm ext}$ is the total external torque applied on the star, and $I$ is now the moment of inertia of the non-rotating star. In the application we consider below there will be two contributions to the total external torque, namely an accretion-induced torque and the torque on the star due to tidal interactions. In terms of sums over modes, Eq.(\ref{dotOmegafinal}) can be rewritten as
\bea
\dot{\Omega} &=& \frac{\mathcal{T}_{\rm ext}}{I} - \frac{1}{I}\frac{d}{dt} \sum_{A,B} c_A^\ast c_B^{\,} \left\langle \hat{\bmath{\Omega}}\times \bmath{\xi}_A,\left[\frac{1}{2}\bmath{B} - i\omega_B\right]\bmath{\xi}_B \right\rangle  \nn \\
&\equiv& \frac{\mathcal{T}_{\rm ext}}{I} - \frac{1}{I}\frac{d}{dt} \sum_{A,B} c_A^\ast c_B^{\,} \Omega \mathcal{K}_{AB},
\label{dotOmegamodes}
\eea
where (\ref{phasespaceexpansion}) has been used. Since $\bmath{\xi}_A \propto e^{im_A \phi}$, the only terms contributing to the above double sum have $m_A = m_B$, i.e. $\mathcal{K}_{AB} \propto \delta_{m_Am_B}$. For modes that do not resonate, we may use approximate solution (\ref{cAapproxI}) for the mode amplitudes to split the double sum of (\ref{dotOmegamodes}) in two pieces, one that includes terms where either mode $A$ or mode $B$ resonates and another involving modes that do not resonate. We write this split as follows
\bea
\dot{\Omega} &=& \frac{\mathcal{T}_{\rm ext}}{I} - \frac{1}{I}\frac{d}{dt} {\sum_{A,B}}^{({\rm r})} c_A^\ast c_B^{\,} \Omega \mathcal{K}_{AB} \nn \\ 
&& - \frac{1}{I}\frac{d}{dt} {\sum_{A,B}}^{({\rm nr})} \frac{\Omega \mathcal{F}_A^\ast\mathcal{F}_B }{(m_A\dot{u} - \omega_A)(m_A\dot{u} - \omega_B)}\mathcal{K}_{AB}.
\eea
Since all terms in the sum over non-resonant modes depend on time only through the star's spin frequency\footnote{Orbital frequency also depends on time but in this paper we will keep orbital parameters fixed for the purpose of computing the time evolution of the spin frequency. We motivate this approximation in the next subsection.}, we may replace the operator $d/dt$ acting on each term of this sum by $\Delta_{AB}\dot{\Omega}/\Omega$, with $\Delta_{AB}$ being of order unity. This leads to the following expression
\be\label{OmegadotwithIeff_app}
\dot{\Omega} = \frac{1}{I_{\rm eff}} \left\{\mathcal{T}_{\rm ext} - \frac{d}{dt} {\sum_{A,B}}^{({\rm r})} c_A^\ast c_B^{\,} \Omega \mathcal{K}_{AB} \right\},
\ee
where the effective moment of inertia $I_{\rm eff}$ is given by
\bea
I_{\rm eff} &=& I\left\{1 + \frac{1}{I}{\sum_{A,B}}^{({\rm nr})}\frac{\Delta_{AB} \mathcal{F}_A^\ast\mathcal{F}_B }{(m_A\dot{u} - \omega_A)(m_B\dot{u} - \omega_B)} \mathcal{K}_{AB}\right\}. \nn \\ && \, \label{Ieff_app}
\eea

Although we are unable to prove that the sum over all non-resonant modes is small compared to unity, we suspect it is the case since a typical driving amplitude $\mathcal{F}_A$ scales as $R/d$ to some power $p \geq 2$, where $R$ is the accretor's radius and $d$ orbital separation. Even though the binary is quite compact (i.e. $R/d$ is not that small, maybe of order $\sim 0.2$), the leading term in the sum over non-resonant modes, which scales as $(R/d)^4$, is of order $10^{-2} - 10^{-3}$. It thus appears unlikely to us that the sum over all non-resonant modes may contribute to more than a few percent to the effective moment of inertia $I_{\rm eff}$ defined in (\ref{Ieff_app}).

\section{Brief overview of generalized {\it r}-modes}\label{modeidentities}

Here we briefly summarize the properties of generalized {\it r}-modes needed for this paper. This appendix relies heavily on the work of \cite{jb1,jb2,lindblom}. Note however that we use the frequency sign convention of \cite{Schenk} instead of the frequency convention of \cite{jb1,jb2,lindblom}. For $l \geq 2$ the Lagrangian displacement of a given mode $\bmath{\xi}_A$ can be written as
\be\label{modefunction}
(\bmath{\xi}_A)^i = \frac{\psi_A}{4\Omega^2}\tilde{Q}^{ij}_A\partial_j \delta U_A,
\ee
where $\psi_A$ is a normalization constant fixed below and where the hydrodynamic potential $\delta U_A$ is given by 
\be\label{deltaU}
\delta U_A = \beta_A \frac{P^m_l(\xi)}{P^m_l(\xi_0)}P^m_l(\tilde{\mu})e^{im_A\phi},
\ee
where $\beta_A$ is a constant and where $(\xi,\tilde{\mu},\phi)$ are a bispheroidal coordinate system defined below. We will henceforth drop all $A$ subscripts for notational convenience since no ambiguity occurs in what follows. In cartesian coordinates, the tensor $\tilde{Q}^{ij}$ is given by
\be\label{Qtensor}
\tilde{Q}^{ij} = \frac{1}{w^2 - 1}\left[\delta^{ij} - \frac{1}{w^2}\delta^{iz}\delta^{jz} + \frac{i}{w}\epsilon^{ijz}\right],
\ee
where $w = \omega / 2\Omega$ and $\epsilon^{ijz} = \delta^{ix}\delta^{jy} - \delta^{iy}\delta^{jx}$. The bispheroidal coordinates $(\xi,\tilde{\mu},\phi)$, essential to write a separated solution for $\delta U$, are explained in detail in \cite{lindblom}. They are related to cartesian coordinates by the following relations
\bea
&& x = b(1-\xi^2)^{1/2}(1-\tilde{\mu}^2)^{1/2}\cos\phi, \\
&& y = b(1-\xi^2)^{1/2}(1-\tilde{\mu}^2)^{1/2}\sin\phi, \\
&& z = b\left(\frac{\sqrt{1 - w^2}}{w}\right)\xi\tilde{\mu},
\eea
where 
\be
b^2 = \frac{a^2}{1-w^2}\left[(1+\zeta_0^2) - w^2\right],
\ee
the parameters $a$ and $\zeta_0$ being the standard parameters describing the background spheroid, namely $a$ is the focal radius and $\zeta_0$ is related to the spheroid's eccentricity by $e^2 = 1/(1+\zeta_0^2)$. Note also that bispheroidal coordinates depend on normal mode frequency. The volume of the star is covered by $\xi_0 \leq \xi \leq 1$, $-\xi_0 \leq \tilde{\mu} \leq \xi_0$, where 
\be
\xi_0^2 = \frac{a^2\zeta_0^2}{b^2}\frac{w^2}{1-w^2}.
\ee
We use the convention of \cite{jb1} for mode normalization, namely a mode of unit amplitude has (rotating frame) energy equal to the fixed value $MR^2\Omega^2$. In the slow rotation approximation one can show (\cite{jbthesis}) that the normalization constant $\psi$ is 
\be\label{psialpha}
\psi = \frac{2M_1}{R_1}\frac{\tilde{\Omega}^2}{\beta} \left(\frac{1-w^2}{3} \right)^{1/2}\left[\frac{(l-m)!}{(l+m)!}\frac{(2l+1)}{l(l+1)}\right]^{1/2},
\ee
where $\tilde{\Omega} = \Omega / \Omega_{\rm break-up} \equiv \Omega / \sqrt{M_1/R_1^3}$. Lastly we give a formula for the density perturbation of a normalized mode
\be
\delta \rho = \psi \rho^2 \delta(p) \Big[\delta U + \delta \Phi\Big],
\ee
where $\delta(p)$ is a delta function with the background pressure as its argument and $\delta \Phi$ is the gravitational potential perturbation of the mode. The density perturbation is non-zero only on the surface $\Sigma$ of the star. On this surface the function $\delta \Phi$ is equal to 
\be\label{deltaPhisigma}
\delta \Phi|_\Sigma = \beta D(\zeta_0) P^m_l(\mu) e^{im\phi},
\ee
where $D(\zeta_0)$ is a known function of $\zeta_0$ [cf. Eqs.(7.1) and (7.4) of \cite{lindblom}]. It is also important to note that on the surface of the star, we have $\delta U|_\Sigma = \beta P^m_l(\mu) e^{im\phi}$ (\cite{lindblom}). In the slowly rotating limit, one can show
\be\label{Dzeta0}
D(\zeta_0) = \frac{3}{2(l-1)}.
\ee
The coordinate $\mu$ appearing first in (\ref{deltaPhisigma}) is part of the spheroidal coordinate system defined by the relations
\bea
&& x = a(1+\zeta^2)^{1/2}(1-{\mu}^2)^{1/2}\cos\phi, \\
&& y = a(1+\zeta^2)^{1/2}(1-{\mu}^2)^{1/2}\sin\phi, \\
&& z = a\zeta{\mu}.
\eea
In the slowly rotating limit, the spheroidal coordinate system becomes the usual spherical coordinate system $(r,\theta,\phi)$ on the surface of the star. In this approximation we then have
\bea
\delta \rho &=& \psi \beta \frac{2l+1}{2(l-1)}\rho^2 \delta(p) P^m_l(\cos\theta)e^{im\phi} \nn \\
&=& \frac{M}{R^2}\frac{\tilde{\Omega}^2}{4}\frac{2l+1}{l-1}\left[\frac{3(1-w^2)}{\pi l (l+1)}\right]^{1/2}\delta(r-R) Y_{lm}(\theta,\phi),\nn\\ \label{deltarhoslowrot}
\eea
where we made use of (\ref{psialpha}), (\ref{Dzeta0}) and the slow rotation limit of the pressure distribution of an incompressible fluid spheroid, which is
\be
p = \frac{2\pi \rho^2}{3}(R^2 - r^2).
\ee

\section{Tidal coupling torque}\label{tidalcouplingtorque}

Here we derive the rate of change of orbital angular momentum induced by the excitation of normal modes in one of the stars of the binary. This result is well known but we include the derivation here in the context of perturbations of rotating stars for completeness. Consider a rotating star perturbed by the following Lagragian fluid displacement vector field
\be
\bmath{\xi}(\bmath{x},t) = \sum_{A} c_A(t) \bmath{\xi}_A(\bmath{x})
\ee

The net force $\bmath{F}_{\rm ext}$ on the perturbed star due to the companion's tidal field is given by
\be
\bmath{F}_{\rm ext} = \int dM \bmath{a}_{\rm ext}(\bmath{y},t),
\ee
where $dM$ is the mass of a fluid element and $\bmath{y} = \bmath{x} + \bmath{\xi}$ is the location of that fluid element. Now we have $dM = d^3x\,  \rho(\bmath{x})$, where $\rho(\bmath{x})$ is the background density and $\bmath{x}$ is the background location of the fluid element. To leading order in the perturbation we then get
\bea
\bmath{F}_{\rm ext} &=& \int d^3x \, \rho(\bmath{x}) \bmath{a}_{\rm ext}(\bmath{x}+ \bmath{\xi},t) \nn \\
&=& \int d^3x \, \rho(\bmath{x}) \left\{ \bmath{a}_{\rm ext}(\bmath{x},t) + \bmath{\xi} \cdot \bmath{\nabla} \bmath{a}_{\rm ext}(\bmath{x},t) \right\} \nn \\ 
&\equiv& \bmath{F}^{(0)}_{\rm ext}(t) + \bmath{F}^{(1)}_{\rm ext}[\bmath{\xi}].
\eea
The zeroth order force term describes the binary's motion when no modes are excited. The term depending on $\bmath{\xi}$ can be rewritten as follows
\bea
\bmath{F}^{(1)}_{\rm ext}[\bmath{\xi}] &=& \int d^3x \, \rho(\bmath{x}) \, \xi_i(\bmath{x},t) \partial_i \bmath{a}_{\rm ext}(\bmath{x},t) \nn \\
&=& \sum_A \int d^3x \, \rho(\bmath{x}) \, c_A(t) \xi_{A}^i(\bmath{x}) \partial_i \bmath{a}_{\rm ext}(\bmath{x},t)  \nn \\
&=& \sum_A c_A(t) \int d^3x \, \delta \rho_A(\bmath{x}) \, \bmath{a}_{\rm ext}(\bmath{x},t) .
\eea
The external acceleration is given by
\bea 
\bmath{a}_{\rm ext}(\bmath{x},t) &=& - \bmath{\nabla} \Phi_{\rm ext}\big(|\bmath{x} - \bmath{d}(t)|\big) \nn \\
&=& \bmath{\nabla}^{(\bmath{d})}  \Phi_{\rm ext}\big(|\bmath{x} - \bmath{d}|\big),
\eea
where $\bmath{d}$ is the binary's orbital separation vector and $\bmath{\nabla}^{(\bmath{d})}$ denotes the gradient operator with respect to $\bmath{d}$, keeping the field point $\bmath{x}$ fixed. We then have
\bea 
\bmath{F}^{(1)}_{\rm ext}[\bmath{\xi}] &=& \sum_A c_A \bmath{\nabla}^{(\bmath{d})} \int d^3x \,\, \delta \rho_A \Phi_{\rm ext} \nn \\ 
&=& \sum_A - i c_A b_A \bmath{\nabla}^{(\bmath{d})} f_A^\ast ,
\eea
where (\ref{fAdeltarho}) has been used. If we write $\bmath{d} = d(\cos \phi, \sin\phi)$, the binary's equation of motion (ignoring gravitational radiation reaction and mass transfer) is then
\bea
\ddot{d} - d\dot{\phi}^2 + \frac{M_{\rm tot}}{d^2} = \frac{1}{\mu} \sum_A i c_A b_A \partial_d f_A^\ast  \\
d\ddot{\phi} + 2\dot{d}\dot{\phi} = \frac{1}{\mu d} \sum_A i c_A b_A \partial_\phi f_A^\ast 
\eea
The rate of change of orbital angular momentum due to tidal coupling is then given by
\bea
\dot{J}_{\rm tidal} &=& \sum_A i c_A b_A \partial_\phi f_A^\ast  \nn \\
&=& -\sum_A m_A c_A b_A  f_A^\ast  \nn \\
&=& -\sum_A m_A c_A b_A \Big[\dot{c}^\ast_A - i\omega_A c_A^\ast \Big], \label{tidaltorque}
\eea
where (\ref{fAdef}) has been used. Using mode pair notation, i.e. Eqs.(\ref{modepairnotationA}) to (\ref{modepairnotationD}), (\ref{balphaA})-(\ref{balphaB}) and (\ref{calphaA})-(\ref{calphaB}), we may further simplify this result to finally arrive at
\be\label{tidaltorquefinal}
\dot{J}_{\rm tidal} = - \sum_{\alpha} m_\alpha b_{\alpha} \frac{d}{dt}|c_\alpha|^2.
\ee

\section{Computation of tidal coupling and back-reaction parameters}\label{integrals}

In this appendix we gather analytic results for integrals required in section \ref{models}. 

\subsection{Mode amplitude driving term}

The first quantity we compute is the driving term $\mathcal{F}_\alpha$ appearing in Eq.(\ref{modeampEOMII}). To compute this quantity we need
\be
\langle \bmath{\xi}_\alpha , -\bmath{\nabla} \Phi_{\rm ext} \rangle = -\int d^3x \, \delta \rho^\ast_\alpha \Phi_{\rm ext}.
\ee    
Using (\ref{Phiextexpansion}) and (\ref{deltarhoslowrot}) we immediately obtain 
\bea
\langle \bmath{\xi}_\alpha , -\bmath{\nabla} \Phi_{\rm ext} \rangle &=& \tilde{\Omega}^2\frac{M_1M_2}{d}\left(\frac{R_1}{d}\right)^l W_{lm} \frac{(2l+1)}{4(l-1)}\nn \\
&& \times\left[\frac{3(1-w_\alpha^2)}{\pi l (l+1)}\right]^{1/2} e^{-im_\alpha u}.
\eea
The normalization convention we choose in Appendix \ref{modeidentities} implies $b_\alpha = M_1R_1^2\Omega / (2w_\alpha)$ (see also Appendix A of \cite{Schenk}) and we get
\bea
\mathcal{F}_\alpha &=& \frac{iw_\alpha \tilde{\Omega}M_2}{M^{1/2}_1 R_1^{3/2}}\left[\frac{R_1}{d}\right]^{l+1} W_{lm} \frac{(2l+1)}{2(l-1)}\sqrt{\frac{3(1-w_\alpha^2)}{\pi l (l+1)}} \nn \\ 
&=& iqw_\alpha \Omega \left[\frac{R_1}{d}\right]^{l+1} W_{lm} \frac{(2l+1)}{2(l-1)}\sqrt{\frac{3(1-w_\alpha^2)}{\pi l (l+1)}},\label{Falphaansw}
\eea
where $q = M_2/M_1$ is the mass ratio.

\subsection{Integrals in the spin frequency evolution equation}

We next compute some of the integrals appearing in Eq.(\ref{dotOmegamodes}). These are more tricky since they cannot be reduced to contributions from the surface of the star alone. It possible to evaluate these integrals analytically for a fluid displacement vector composed of a superposition of generalized {\it r}-modes using a recursion technique (\cite{jbthesis}), but a general closed-form expression for the integrals appearing in Eq.(\ref{dotOmegamodes}) is not available. In the model where a single {\it r}-mode pair is assumed to be dynamically relevant however, the required integral is 
\be
\frac{1}{2}\kappa_\alpha = -i\omega_\alpha\langle \bmath{z}\times\bmath{\xi}_\alpha, \bmath{\xi}_\alpha \rangle + \Omega \langle \bmath{z}\times\bmath{\xi}_\alpha, \bmath{z}\times\bmath{\xi}_\alpha \rangle,
\ee
which we now give in the slow rotation approximation. Using (\ref{modefunction}) and (\ref{Qtensor}) we have, dropping the $\alpha$ and $1$ subscripts below for brevity,
\bea
\frac{1}{2}\kappa &=& \Omega \frac{\psi^2\rho}{16\Omega^4(1-w^2)^2} \int d^3x \left(\delta^{jp} - \frac{1}{w^2}z^jz^p - \frac{i}{w}\epsilon^{jpz}\right)\nn \\
&& \times \Big(X_{jk} - 2iw\,\epsilon_{jkz}\Big)\left(\delta^{kq} - \frac{1}{w^2}z^kz^q + \frac{i}{w}\epsilon^{kqz}\right)\nn \\
&& \times \partial_p\delta U^\ast \partial_q \delta U,
\eea
where $X_{jk} = \delta_{jk} - z_jz_k$. Performing the index contractions gives
\bea
\frac{1}{2}\kappa &=& \Omega \frac{\psi^2\rho}{16\Omega^4(1-w^2)^2} \int d^3x \left[\left(\frac{1-w^2}{w^2} -2\right)X^{pq} \right. \nn \\
&&   - 2iw\, \epsilon^{pqz}\Bigg]\partial_p\delta U^\ast \partial_q \delta U.
\eea
The remaining integrals have been evaluated in \cite{jbthesis} for spheroids of arbitrary ellipticity. In the slow rotation approximation, the results are
\be
\int d^3x \, i\epsilon^{pzq}\partial_p\delta U^\ast \partial_q \delta U = 2\pi \beta^2 \frac{2m}{2l+1}\frac{(l+m)!}{(l-m)!}R
\ee
\bea
\int d^3x \, X^{pq} \partial_p\delta U^\ast \partial_q \delta U =  2\pi \beta^2\frac{1}{2l+1}\frac{(l+m)!}{(l-m)!}\frac{R}{w}\nn \\
 \times \Big[-2m + (1-w^2)(m + l(l+1)w) \Big] .
\eea
Using (\ref{psialpha}) we arrive at
\bea
\kappa = \frac{MR^2\Omega }{4l(l+1)(1-w^2)}\left\{4mw + \frac{1}{w}\left(\frac{1-w^2}{w^2} - 2\right)\right. \nn \\
\left.\times \Big[-2m + (1-w^2)(m + l(l+1)w)\Big]\right\}. \label{taures}
\eea
We then obtain $\bar{\nu}$ by combining (\ref{taures}) and (\ref{barnudef}). In table \ref{table:numericalvalues}, we report numerical values for $\lambda$, $|\mathcal{F}|/\Omega$ and $\bar{\nu}/\Omega$ for all {\it r}-modes with $m > 0$, $l \leq 6$ and $l+m$ even. In evaluating $\mathcal{F}$ and $\bar{\nu}$ we use $I = 0.4 MR^2$ for the moment of inertia of the accretor, $R/d = 0.2$ for the tidal ratio and $q=0.25$ for the mass ratio. It is easy to rescale the numbers given in table \ref{table:numericalvalues} for other values of $I$, $R/d$ and $q$.

\begin{table}
\caption{\label{table:numericalvalues} Numerical values for the parameter $\lambda = m + 2w$, the resonant spin frequency $\Omega_{\rm res}$, the tidal coupling $|\mathcal{F}|$ and the back-reaction parameter $\bar{\nu}$ for generalized {\it r}-modes with $l+m$ even, up to $l=6$.}
\begin{tabular}{lcccc}
\hline
$(l,m,w)$ & $\lambda$ & $\Omega^{\rm res}/\omega_{\rm orb}$ & $|\mathcal{F}|/\Omega$ & $\bar{\nu}/\Omega$ \\ 
\hline \\ 
$3,1,-0.755$ & -0.510 & - &5.6719e-05 & 1.8537 \\ 
$3,1,0.0883$& 1.177 &0.850 &1.0076e-05 & -25.052\\ 
$4,2,-0.616^\ast$ & 0.768 & 2.604 &5.9429e-06 & 7.5596 \\ 
$4,2,0.116$ & 2.232 & 0.896 &1.4105e-06 & -38.004 \\ 
$5,1,-0.903$ & -0.806 & - &5.5717e-07 & 0.84994 \\ 
$5,1,-0.5228$& -0.046 & - &6.4004e-07 & 5.1797 \\ 
$5,1,0.0341$ & 1.068 & 0.936 &4.8944e-08 & -62.488 \\ 
$5,1,0.5917$ & 2.183 & 0.458 &6.8504e-07 & -4.2286 \\ 
$5,3,-0.5266^\ast$& 1.947 & 1.541 &6.9443e-07 & 14.566 \\ 
$5,3,0.1266$ & 3.253 & 0.922 &1.9480e-07 & -52.365 \\ 
$6,2,-0.8217^\ast$ & 0.357 & 5.608 & 9.2482e-08 & 4.551 \\ 
$6,2,-0.4421^\ast$ & 1.116 & 1.792 &7.8312e-08 & 12.779 \\ 
$6,2,0.0509 $ & 2.102 & 0.952 &1.0039e-08 & -83.960 \\ 
$6,2,0.5463 $ & 3.093 & 0.647 & 9.0364e-08 & -8.8759 \\ 
$6,4,-0.464^\ast $ & 3.072 &1.302 & 8.8912e-08 & 22.769 \\ 
$6,4,0.1306$ & 4.261 & 0.939 &2.8022e-08 & -67.972 \\ 
\hline
\end{tabular}

\medskip

The following parameters were used for the moment of inertia of the accretor, the tidal ratio and the mass ratio respectively:  $I = 0.4 MR^2$, $R/d = 0.2$ and $q=0.25$. It is straightforward to rescale the numbers for other values of mass and tidal ratios. The mode quantum numbers marked with a ${\,}^\ast$ are candidates for tidal synchronization. 

\end{table}

\end{document}